\documentclass[aps,showkeys,showpacs,tightenlines,nofootinbib,prd]{revtex4}
\usepackage{amsmath,amscd,amssymb}
\usepackage{color,epsfig}
\usepackage{xfrac}
\usepackage{latexsym}
\usepackage{mathrsfs}
\usepackage[caption=false]{subfig}
\usepackage{graphicx}
\usepackage{ulem}
\usepackage{bbm}
\DeclareMathOperator\sech{sech}

\begin{document}

\title{Scattering of Kinks in Noncanonical sine-Gordon Model}

\author{I. Takyi$^{a)}$, B. Barnes$^{a)}$, H. M. Tornyeviadzi$^{b)}$ and J. Ackora-Prah$^{a)}$}    
 
\affiliation{$^{a)}$Department of Mathematics, Kwame Nkrumah University of Science and Technology, Private Mail Bag, Kumasi, Ghana\\
$^{b)}$Smart Water Lab, Faculty of Engineering Sciences, Norwegian University of Science and Technology, Alesund-Norway\\
Email: ishmael.takyi@knust.edu.gh}

\begin{abstract}
In this paper, we numerically study the scattering of kinks in the noncanonical sine-Gordon model using Fourier spectral methods. The model depends on two free parameters, which control the localized inner structure in the energy density and the characteristics of the scattering potential. It has been conjectured that the kink solutions in the noncanonical model possess inner structures in their energy density, and the presence of these yields bound states and resonance structures for some relative velocities between the kink and the antikink. In the numerical study, we observed that the classical kink mass decreases monotonically as the free parameters vary, and yields bion-formations and long-lived oscillations in the scattering of the kink-antikink system.
\end{abstract}


\maketitle

\section{Introduction}

Kink solitons in $1+1$-dimensional space are an example of topological defects in classical field theory. These kinks are localized with non-dispersive energy density~\cite{Rajaraman:1982is} and behave like classical pointlike particles when subjected to an external force. They are used in many branches of physics because of their stable nature against dispersive effects. For example; they are applied in cosmology for studying the fractal structure of the cosmic domain walls~\cite{Vachaspati:2006zz,Vilenkin:2000jqa}, in condensed matter physics, they are used to study Bose-Einstein condensates~\cite{Kevre2008} and for studying domain walls in ferromagnets~\cite{Ivanov:1992aa} and ferroelectrics~\cite{bishop1980solitons}, in particle physics, they are used as a model of hadrons~\cite{Weigel:2008zz,Weigel:2021pbr,Takyi:2019ahv}.

The scattering of kink and antikink in canonical field models has been of interest to the scientific community, especially mathematical physicists, because of their physical significance, as mentioned earlier. It was observed numerically by the authors in Refs.~\cite{Campbell:1986mg,Campbell:1983xu,Belova:1997bq,Anninos:1991un,Ablowitz:1979a,Moshir:1981ja,Goodman:2005ja,Goodman:2007e} structural patterns of the scattering of kink and antikink in the canonical models. These structures occur by choosing specific initial conditions such that the kink and antikink are well separated. Once boosted with a prescribed velocity, the kink and antikink approach each other and interact. The prescribed velocity is referred to as the relative velocity between the kink and antikink. For example, in the canonical $\phi^{4}$ model, it was discovered that the formation of large-amplitude bound states occurs for initial relative velocities ($v_{\rm in}$) less than a critical velocity ($v_{\rm cr}$) (which separates two distinct classes of solutions). This occurrence is a result of the annihilation of the kink and antikink. Also, for $v_{\rm in} $ less than $ v_{\rm cr}$ an interesting pattern is observed, the so-called resonance windows, which occur when there is an exchange of energy between the shape and translational modes of the kinks, while for $v_{\rm in} $ greater than $ v_{\rm cr}$ the kink and antikink after collision reflect from each other. These phenomena have also been observed in higher polynomial canonical models such as the $\phi^{6}$, $\phi^{8}$ and $\phi^{10}$ models~\cite{Gani:2014gxa,Weigel:2013kwa,Dorey:2011yw,Belendryasova:2019eq,Campos:2020ust,Gani:2015cda,Christov:2018ecz,Christov:2018wsa,Manton:2018deu,Bazeia:2018bhf,Gani:2020pio,Christov:2020zhb} with power-law tails; where power-law tails of kinks result in their long-range interaction. Recently, these structural patterns have also been reported in non-polynomial canonical models~\cite{Bazeia:2017rxo,Bazeia:2019xoe,Takyi:2020nkn} with their interesting physical properties.

There are two ways of studying these structural patterns. One of these ways is by solving the equation of motion, a system of second-order partial differential equations, using a numerical scheme. The other method is by making use of analytical approximations. Examples of these approximations are the collective coordinate method~\cite{Gani:2014gxa,Weigel:2013kwa,Christov:2008kk,Takyi:2016tnc} and the Manton method~\cite{Manton2004,Manton:1978gf,Manton:2021ipk,Kevrekidis:2004ga}. These analytical approximations allow one to estimate the force between the kink and antikink. 

In the canonical sine-Gordon model, the bion-formation and resonance phenomena arise by studying the scattering of kink and antikink from defects or impurities. These impurities are added to the equation of motion as perturbative terms~\cite{Goodman:2004ef,Fei:1992aj,Kivshar:1991zz,Zhang:1991ee,Dorey:2021mdh}. In this case, the resonance structures occur when the kink is reflected by the defect. Also, when the kink remains at the defect for a long time, a bound state occurs and the kinks are reflected when they pass by the defect. Apart from the canonical sine-Gordon model, the resonance structures have also been reported in the study of the collision of several kinks of the double sine-Gordon model at one point~\cite{Gani:2017yla,Gani:1998jb,Campbell:1986nu,Malomed:1989gx,Nazifkar:2010aa}.

In this work, we consider the scattering of kinks in the noncanonical sine-Gordon model. It has been observed generally that, in a noncanonical model, as the parameters vary, the energy density of the kink splits from one peak to multi-peaks~\cite{Zhong:2019fub,Bazeia:2007df,Bazeia:2014dva,Zhong:2018tbi}. This observed feature is the so-called ``inner structure" of the kink. We investigate whether the kinks in the noncanonical sine-Gordon model possess inner structures in their energy density. An interesting phenomenon is the production of long-lived static and moving oscillations in the scattering of the kink-antikink system.

This paper is organized as follows: in section \ref{sec:model} we introduce the model and discuss the properties of the corresponding kink solutions. The numerical results of the scattering of the kink and antikink will be presented in section \ref{collision}. We conclude and summarize in section \ref{sec:concl}.

\label{sec:model}
\subsection{The model structure and kink solution}
We consider the noncanonical scalar field in one-space and one -time dimensional space, whose Lagrangian density is given by 
\begin{equation}
	\mathcal{L} = F(\varphi) X - U(\varphi),
	\label{eq:Lang}
\end{equation}
where $\displaystyle X = -\frac{1}{2} \eta^{\mu\nu} \partial_{\mu} \varphi \partial_{\nu} \varphi$ (with $\eta^{\mu\nu}=\mathrm{diag}(-1,1)$) is the kinetic term of the scalar field, $U$ is the scalar potential and $F(\varphi) = \alpha \left(\sin \varphi\right)^{2n}+1$. Here the parameter $n \geq 0$ accounts for the local maxima of the energy density of the kink, where $\alpha >0$ differentiate the model from the canonical case. 
The Euler equation of motion for this model is given by
\begin{equation}
	\frac{\partial F}{\partial \varphi} \left[\left(\frac{\partial \varphi}{\partial x}\right)^{2}-\left(\frac{\partial \varphi}{\partial t}\right)^{2}\right] + 2F \left[ \frac{\partial^{2} \varphi}{\partial x^{2}} - \frac{\partial^{2} \varphi}{\partial t^{2}}\right] - 2 \frac{\partial U}{\partial \varphi} = 0.
	\label{eq:Euler}
\end{equation}
For static field, $\varphi=\varphi(x)$, the field equation simplifies to the nonlinear static wave equation 
\begin{equation}
	\frac{\partial F}{\partial \varphi} \left(\frac{\partial \varphi}{\partial x}\right)^{2} + 2F  \frac{\partial^{2} \varphi}{\partial x^{2}} - 2 \frac{\partial U}{\partial \varphi} = 0.
	\label{eq:static_wave}
\end{equation}
In what follows, we construct the static kink solutions via the superpotential~\cite{Zhong:2019fub,Bazeia:2008tj,Zhong:2014kha}
\begin{equation}
	W(\varphi) = \frac{\partial \varphi}{\partial x},
	\label{eq:superpot}
\end{equation}
which is related to the scalar field potential by integrating Equ. \eqref{eq:static_wave} and assuming the constant of integration to be zero, this yield
\begin{equation}
	U = \frac{1}{2} FW^{2}.
\end{equation}
Taking the superpotential as
\begin{equation}
	W = \sqrt{2}kv_{0} \sqrt{\left(1-\cos \left(\frac{\varphi}{v_{0}}\right) \right)}
\end{equation}
one obtain the sine-Gordon kink 
\begin{equation}
	\varphi_{K}(x) = 4v_{0} \arctan \left[\exp(kx)\right],
	\label{eq:kinksolution}
\end{equation}
where $v_{0}$ represents the vacuum expectation value of $\varphi(x)$ and $\displaystyle \frac{1}{k}$ is the thickness of the kink. The antikink with $\varphi(-\infty)=2 \pi$ and $\varphi(\infty)=0$ is related to the kink by the spatial reflection $x \leftrightarrow -x$. 
In this paper, we set $v_{0}=1=k$. The scalar potential becomes
\begin{equation}
	U(\varphi) = \left(1-\cos \varphi\right)\left[\alpha\left(\sin \varphi\right)^{2n}+1\right].
	\label{eq:canpot}
\end{equation}
\begin{figure}
	\centering
	\subfloat[Scalar potential $U(\varphi)$ cf. equation \eqref{eq:canpot}.]
	{\includegraphics[scale=0.3]{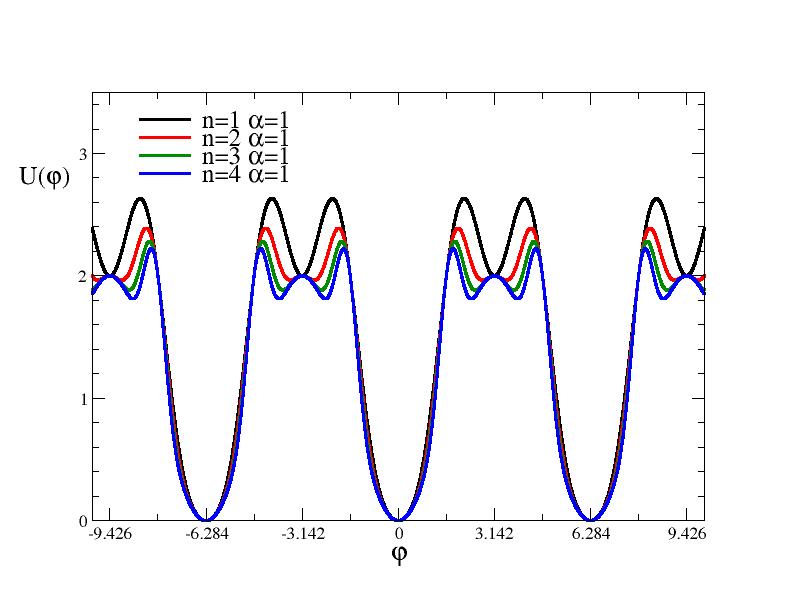}
		\label{fig:scalarpot_a}}
	\quad 
	\subfloat[Energy density of the noncanonical model cf. equation \eqref{eq:energydens}.]
	{\includegraphics[scale=0.3]{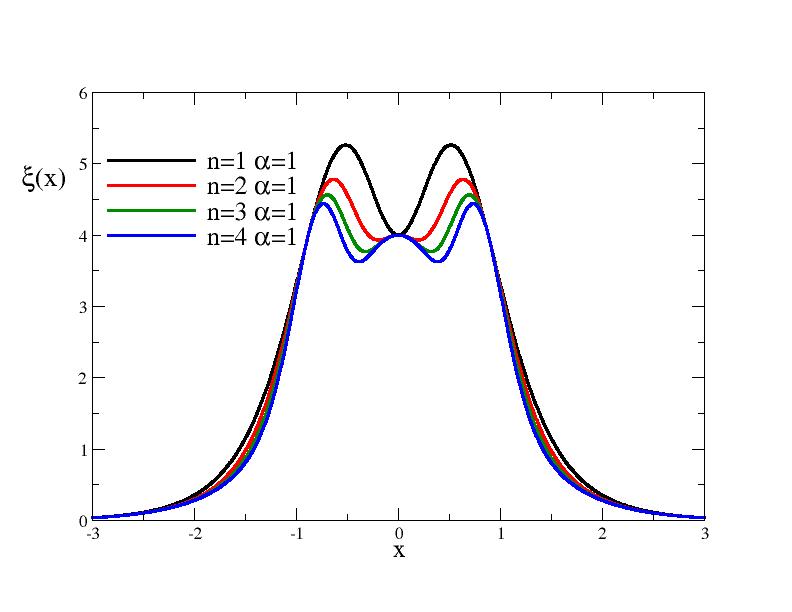} 
		\label{fig:scalarpot_b}}
	\caption{\label{fig:scalarpot} The field potential $U(\varphi)$ and the energy density $\xi (x)$.}
\end{figure}
Taking $\alpha=0$ give the standard sine-Gordon potential, while for $\alpha, \hspace{0.05cm} n>0$ yield potential with vacuum solutions $\varphi_{\rm vac}=2m \pi$, where $m \in \mathbb{Z}$. The kink and antikink solutions interpolate between neighbouring vacua. Thus, there are two discrete symmetries for the above Lagrangian for the potential in this regime. These symmetries are $\varphi \rightarrow -\varphi$ and $\varphi \rightarrow \varphi + 2\pi m$, $m \in \mathbb{Z}$. 
The total energy of the field is given by 
\begin{equation}
	E\left[\varphi\right] = \int_{-\infty}^{\infty} \mathrm{d} x\, \left[ \frac{1}{2} F(\varphi) \left(\frac{\partial \varphi}{\partial t}\right)^{2} + \frac{1}{2} F(\varphi) \left(\frac{\partial \varphi}{\partial x}\right)^{2} + U(\varphi) \right]. \label{eq:totalenergy}
\end{equation} 
We obtain an expression for the energy density of the static kink as
\begin{align}
	\xi(x) & = \frac{1}{2} F(\varphi) \left(\frac{\partial \varphi}{\partial x}\right)^{2}  + U(\varphi) \nonumber \\
	& = 4 \sech^{2} x \left[4^{n} \alpha \sech^{2n}x \tanh^{2n}x+1\right], \label{eq:energydens}
\end{align}
which vanishes at the absolute minima of the potential.

The scalar potentials are depicted in Figure \ref{fig:scalarpot_a}, while the corresponding energy densities are depicted in Figure \ref{fig:scalarpot_b} for $n = 1,2,3,4$ and $\alpha = 1$. We observe that the energy density splits as $n$ increases. For $n=1,2$ and $\alpha=1$ the energy density has two peaks. Three peaks of the energy density are observed for $n = 3,4$ and $\alpha = 1$. The first five values of $n$ for $\alpha \neq 0$ for the classical energy (mass) cf. Equ. \eqref{eq:totalenergy} are shown in Table\ref{t1} which decrease monotonically as $n$ increases.
\begin{table}
	\caption{\label{t1} Classical energy for some values of $n$.}	
	\centerline{
		\begin{tabular}{l  l }
			$n$ & $E_{\rm cl}$ \\
			\hline
			$0$   & $8 \alpha + 8$ \vspace{0.3cm} \\ 
			$1$   & $ \displaystyle \frac{64}{15} \alpha + 8 $ \vspace{0.3cm} \\ 
			$2$   & $ \displaystyle \frac{1024}{315} \alpha + 8 $ \vspace{0.3cm} \\  
			$3$   & $ \displaystyle \frac{8192}{3003} \alpha + 8 $ \vspace{0.3cm} \\ 
			$4$   & $ \displaystyle \frac{262144}{109395} \alpha + 8 $ \vspace{0.2cm}\\  \hline  
	\end{tabular}}
\end{table}
\begin{figure}
	\centerline{
		\includegraphics[width=14cm,height=5cm]{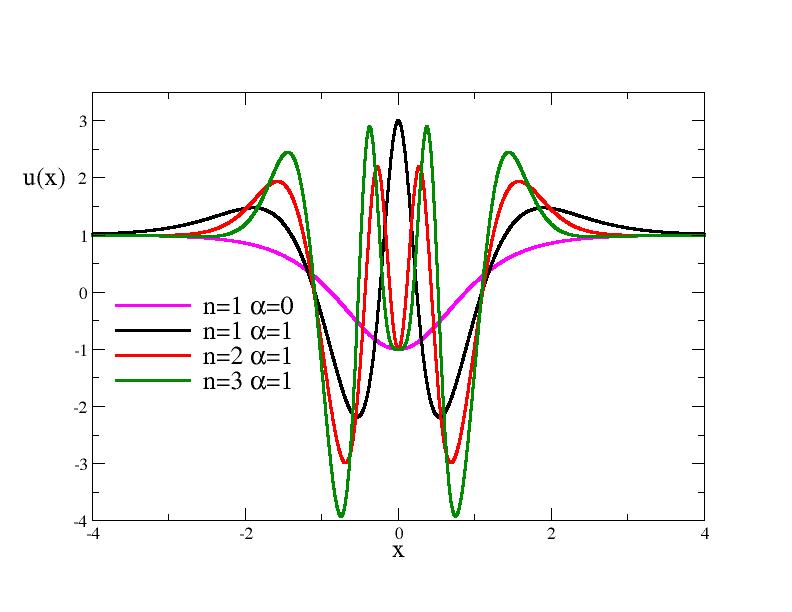}}
	\caption{The scattering potential $u(x)$.}
	\label{fig:scatpot}
\end{figure}
\begin{figure}
	\centering 
	\subfloat[The translational zero modes ]{\includegraphics[scale=0.3]{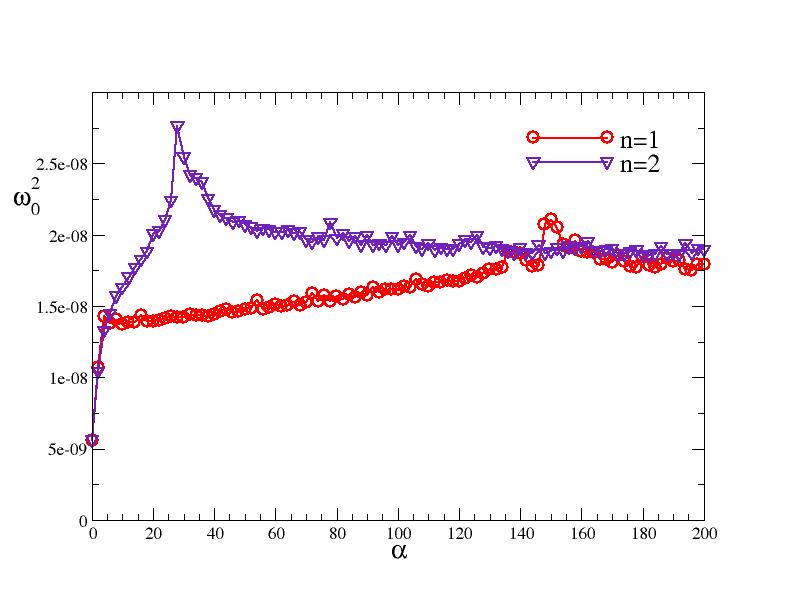} 
		\label{fig:eigenvalue_a}}
	\quad
	\subfloat[The internal shape modes]{\includegraphics[scale=0.3]{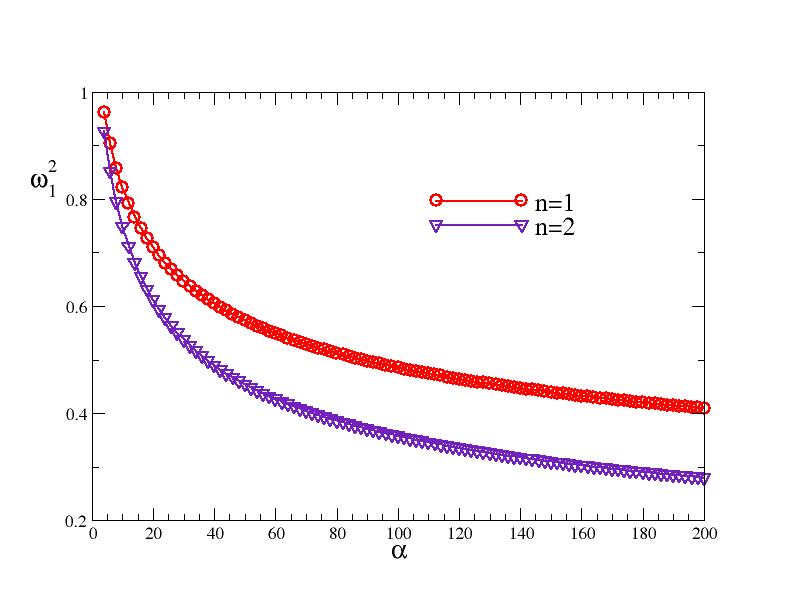}
		\label{fig:eigenvalue_b}}
	\caption{\label{fig:eigenvalue} The eigenvalues (modes) of equation \eqref{eq:wave_equ} for $\alpha \in (0:2:200)$ and $n=1,2$.}
\end{figure}

\subsection{Linear spectrum}
In what follows, we analyse the excitation spectrum of the static kink by considering the fluctuation modes around the kink, $\eta(x,t)=e^{i \omega t} \eta(x)$, such that $\varphi(x,t)= \varphi_{K}(x) + \eta(x,t)$. Substituting this into Equ. \eqref{eq:Euler} and collecting terms linear in $\eta$ with $U=FU_{SG}$, where $U_{SG}=1-\cos \varphi$, gives
\begin{align}
	\frac{\partial^{2}F}{\partial \varphi_{K}^{2}}\varphi_{K}^{\prime \hspace{0.02cm}2}\eta + 2\frac{\partial F}{\partial \varphi_{K}} \varphi_{K}^{\prime}\eta^{\prime} + 2 F\left(\eta^{\prime\prime} - \ddot{\eta} \right) + 2\frac{\partial F}{\partial \varphi_{K}} \varphi_{K}^{\prime\prime} \eta - 2\left( \frac{\partial^{2} U}{\partial \varphi_{K}^{2}} + 2 \frac{\partial F}{\partial \varphi_{K}} \frac{\partial U_{SG}}{\partial \varphi_{K}}\right) \eta =0,
\end{align}
where the primes and dots denote derivatives with respect to the space and time variables respectively. Using the separation ansatz $\eta(x,t)=e^{i \omega t} \eta(x)$, the above equation simplifies to  
\begin{equation}
	\eta^{\prime\prime}= - \omega^{2} \eta - \frac{1}{F} \frac{\partial F}{\partial \varphi_{K}} \varphi_{K}^{\prime} \eta^{\prime} + \left[\frac{\partial^{2} U_{SG}}{\partial \varphi_{K}^{2}} + \frac{1}{F}\frac{\partial F}{\partial \varphi_{K}}\varphi_{K}^{\prime \prime} \right]
\end{equation}
which is not the standard form for the potential scattering because of the $\eta^{\prime}$ term. Parametrizing $\eta=G\xi$, where $\displaystyle G= \frac{1}{\sqrt{F}}$ yields the standard form of the potential scattering as 
\begin{equation}
	\left[-\frac{\mathrm{d}^{2}}{\mathrm{d}x^{2}} + u(x)\right] \xi(x) = \omega^{2} \xi(x),
	\label{eq:wave_equ}
\end{equation}
where 
\begin{align}
	u(x)= \frac{1}{F} \frac{\partial^{2} U}{\partial \varphi_{K}^{2} } + \frac{3}{2} \frac{1}{F} \frac{\partial F}{\partial \varphi_{K}} \varphi_{K}^{\prime \prime} - \frac{1}{4} \left(\frac{1}{F} \frac{\partial F}{\partial \varphi_{K}} \varphi_{K}^{\prime}\right)^{2}
	\label{eq:scat_pot}
\end{align}
is the scattering potential, whose explicit expression is obtained by substituting the kink profile $\varphi_{K}$, the scalar potential $U(\varphi_{K})$ and $F(\varphi_{K})$. Figure \ref{fig:scatpot} shows the scattering potential for $n=1,2,3$ and $\alpha = 1$. The special case for $\alpha=0$ when $n=1$ with a global minimum at $x=0$ is also shown. The figure shows that for $\alpha >0$ for all $n$ there is the possibility of occurrence of bound states. For $\alpha=1$ and $n=1$ the point $x=0$ turns to be a local maximum with two minima in the potential, while for $n=2,3$ and $\alpha=1$ the point $x=0$ turns to be a local minimum with three minima in the potential. The scattering potential is symmetric and the fluctuation masses around the vacua as $x \rightarrow \pm \infty$ remain the same and is equal to $1$. 

We solve the eigenvalue problem by making use of the shooting method. This is done by integrating Equ. \eqref{eq:wave_equ} using the asymptotic behavior of its solutions at $x=\pm \infty$. Starting at a large negative $x$ value, we obtain the `left' solution and obtain a `right' solution at a large positive $x$ value. The two solutions are then matched at some arbitrary matching point $x_{m}$. We chose $x_{m}$ close to the spatial origin to avoid technical issues when $u(x)$ is an even function of $x$. The eigenvalues of $\omega^{2}$ are those values at which the Wronskian of the `left' solution and `right' solution at the matching point $x_{m}$ turns to zero. The results are shown in Figure \ref{fig:eigenvalue} for $\alpha \in \left[0:2:200\right]$ and $n=1,2$. We record an internal shape mode only at $\alpha \geq 4$ cf. Figure \ref{fig:eigenvalue_b} , below this value only translational zero modes (see Figure \ref{fig:eigenvalue_a}) are observed.  
\begin{figure}
	\centering
	\subfloat[$v_{\rm in}=0.501$ for $n=1$ and $\alpha = 1$.]
	{\includegraphics[scale=0.1]{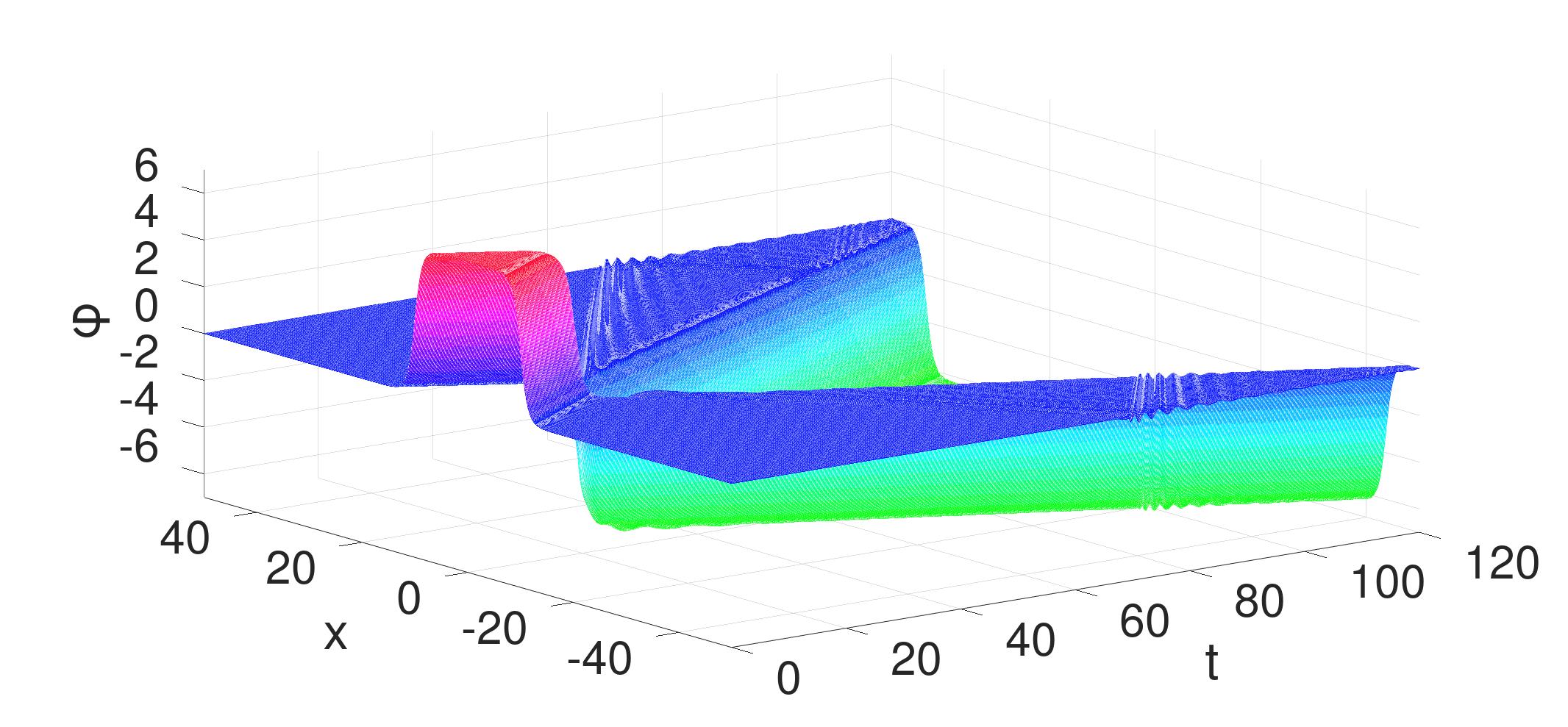} \label{fig:reflection_a}}
	\quad
	\subfloat[$v_{\rm in }=0.4$ for $n=2$ and $\alpha = 1$.]
	{\includegraphics[scale=0.1]{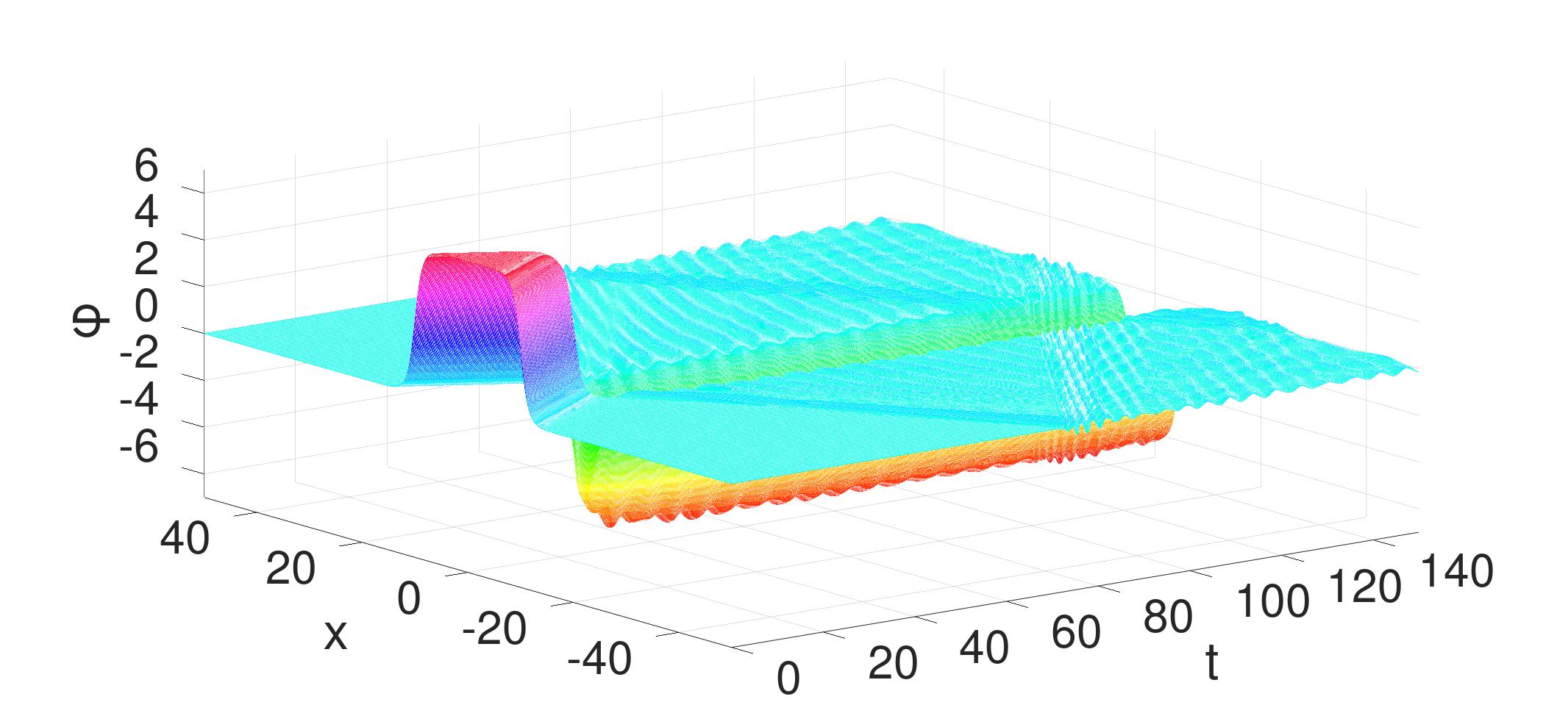} \label{fig:reflection_b}}
	\quad
	\subfloat[$v_{\rm in }=0.601$ for $n=2$ and $\alpha=5$.]
	{\includegraphics[scale=0.1]{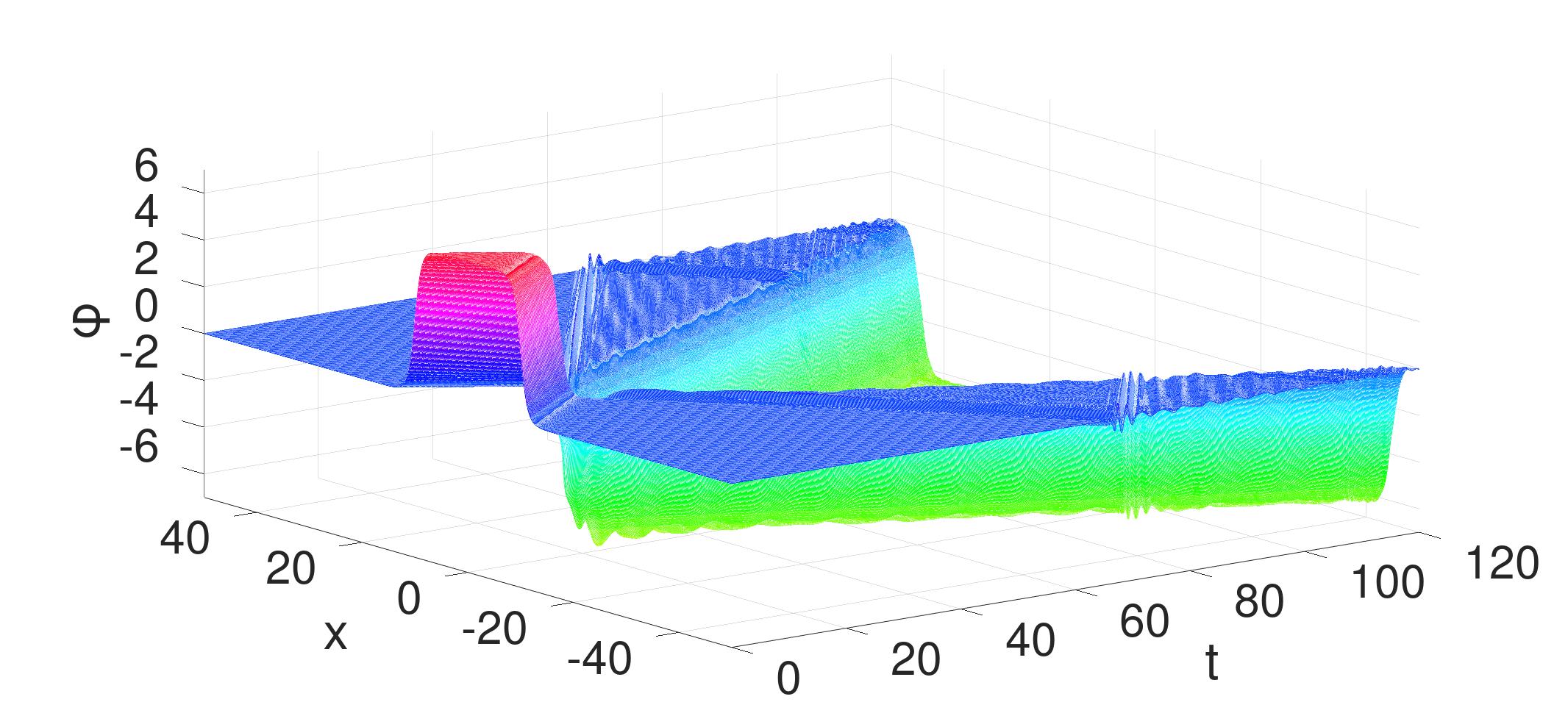}\label{fig:reflection_c}}
	\caption{\label{fig:reflection} Annihilation of the kink and antikink pair to a long-lived oscillating state}
\end{figure}

\section{Numerical results}
\label{collision}
\begin{figure}
	\centering
	\subfloat[$v_{\rm in}=0.361$ for $n=1$ and $\alpha = 1$.]
	{\includegraphics[scale=0.1]{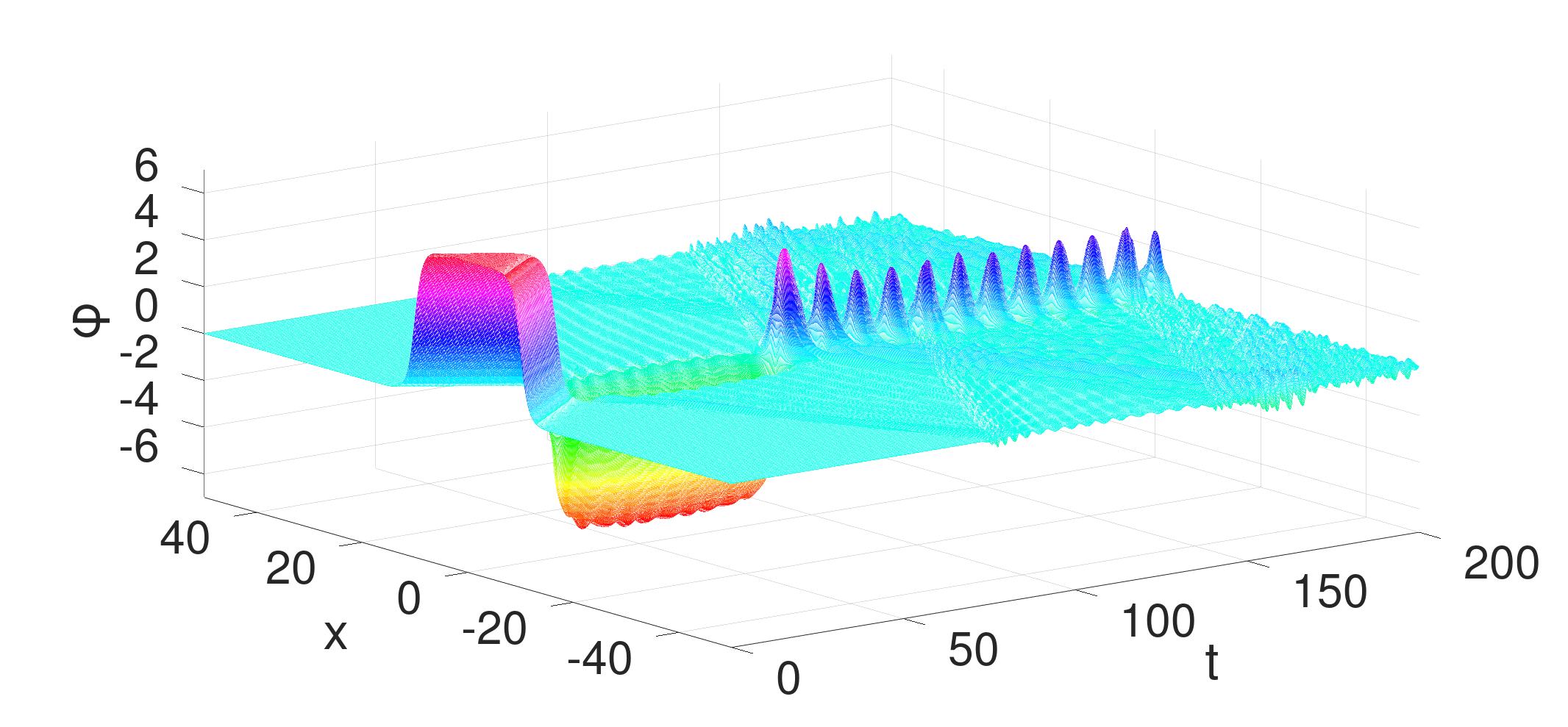}\label{fig:AnnCenBion_a}}
	\quad 
	\subfloat[$v_{\rm in }=0.3621$ for $n=1$ and $\alpha = 1$.]
	{\includegraphics[scale=0.1]{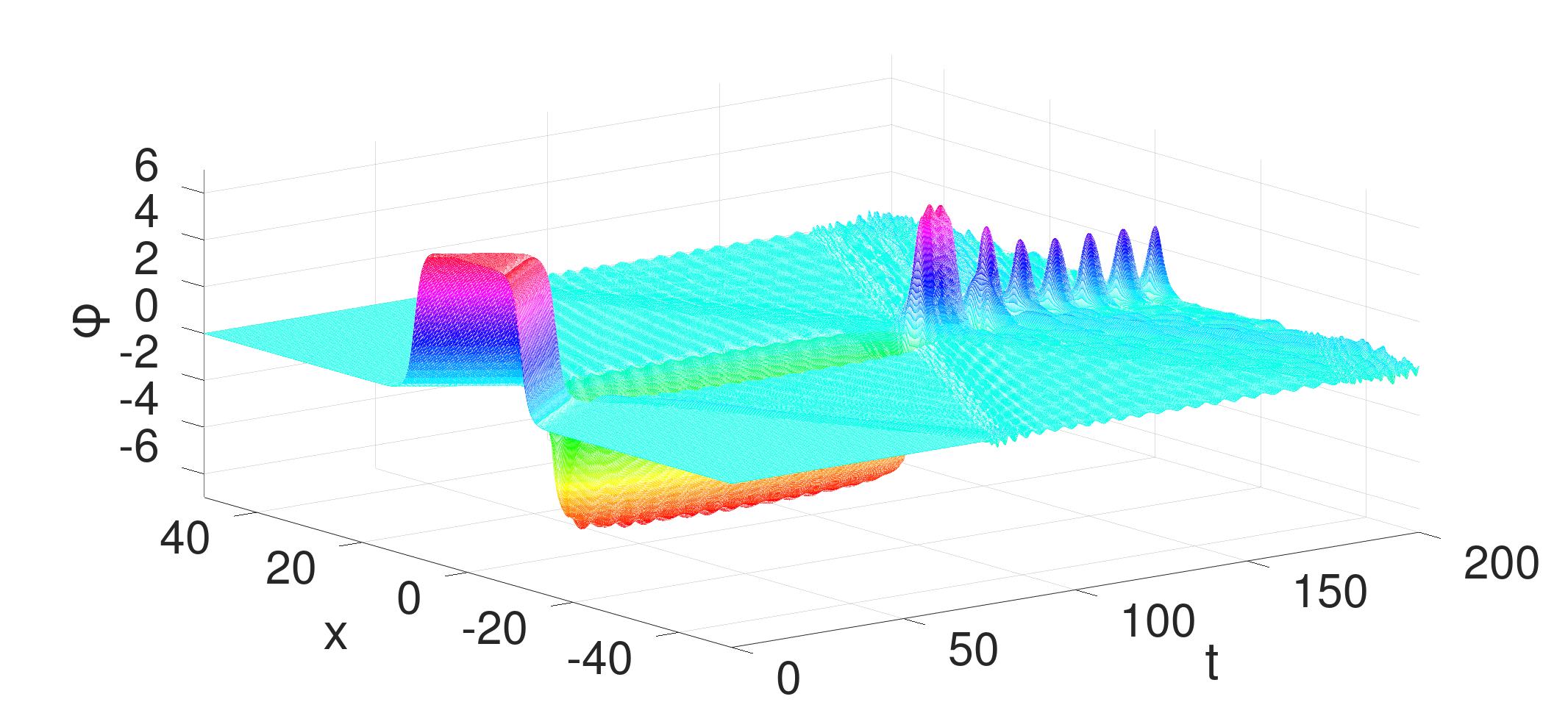}\label{fig:AnnCenBion_b}}
	\quad 
	\subfloat[$v_{\rm in }=0.3961$ for $n=2$ and $\alpha=1$.]
	{\includegraphics[scale=0.1]{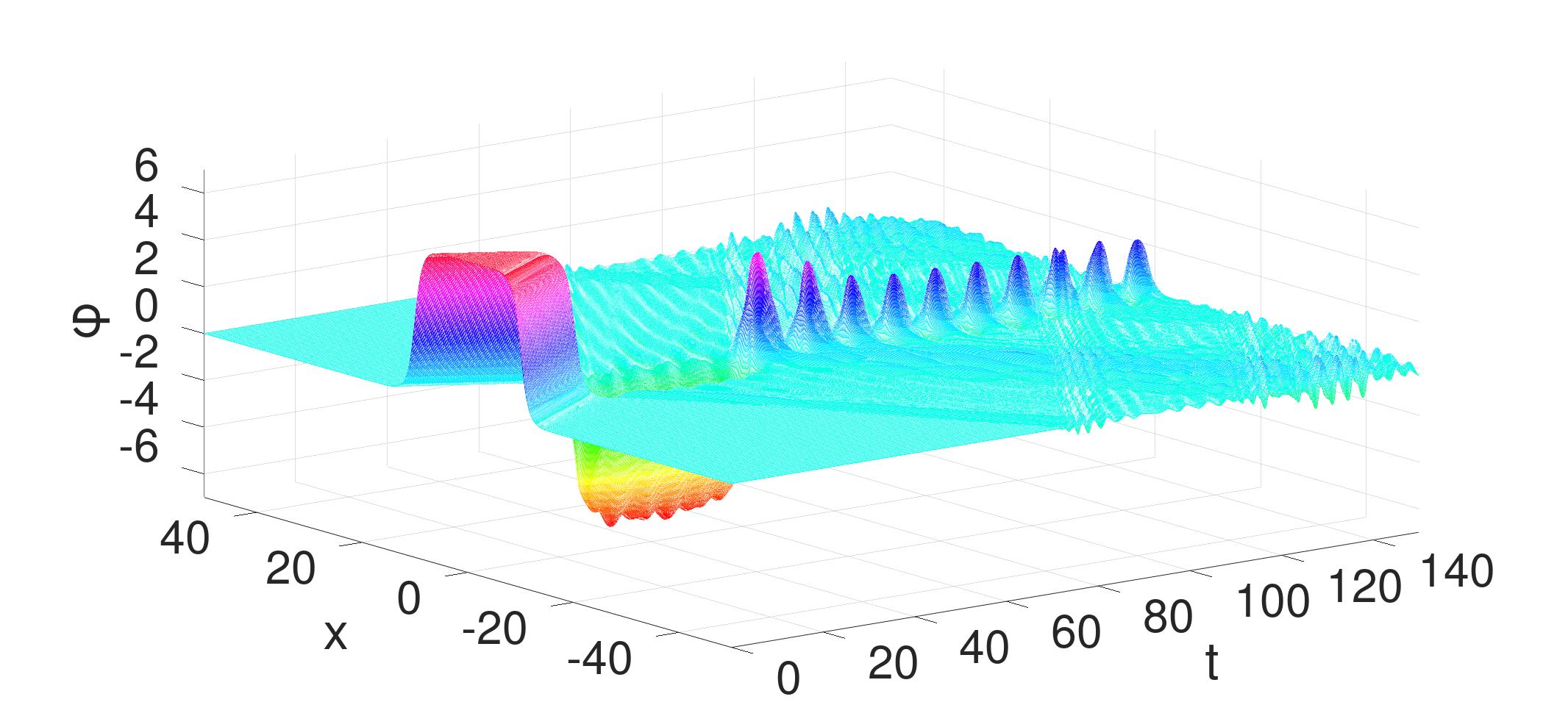}\label{fig:AnnCenBion_c}}
	\caption{\label{fig:AnnCenBion}An annihilation of a kink and antikink pair to a centrally-located bion.}
\end{figure}
The presence of the excitation modes in the kink means that, at a certain instant, there is a transfer of the kinetic energy of the moving kink and antikink to the excitation modes such that the kink and antikink are unable to overcome their attractive potential, resulting in a trapped state; but for some initial relative velocities, the internal shape mode is destroyed and its energy is transferred into the translational zero modes and the kink and antikink propagate almost separately. These are the so-called resonance windows. 
We investigate the interaction of the kink and antikink systems by solving the dynamical system equation numerically. This is done by taking the kink and antikink solutions as the initial condition, wherein initially the kink and antikink are widely separated whilst propagating towards each other. We make use of the superposition ansatz
\begin{figure}
	\centering
	\subfloat[$v_{\rm in}=0.101$ for $n=1$ and $\alpha = 1$.]
	{\includegraphics[scale=0.1]{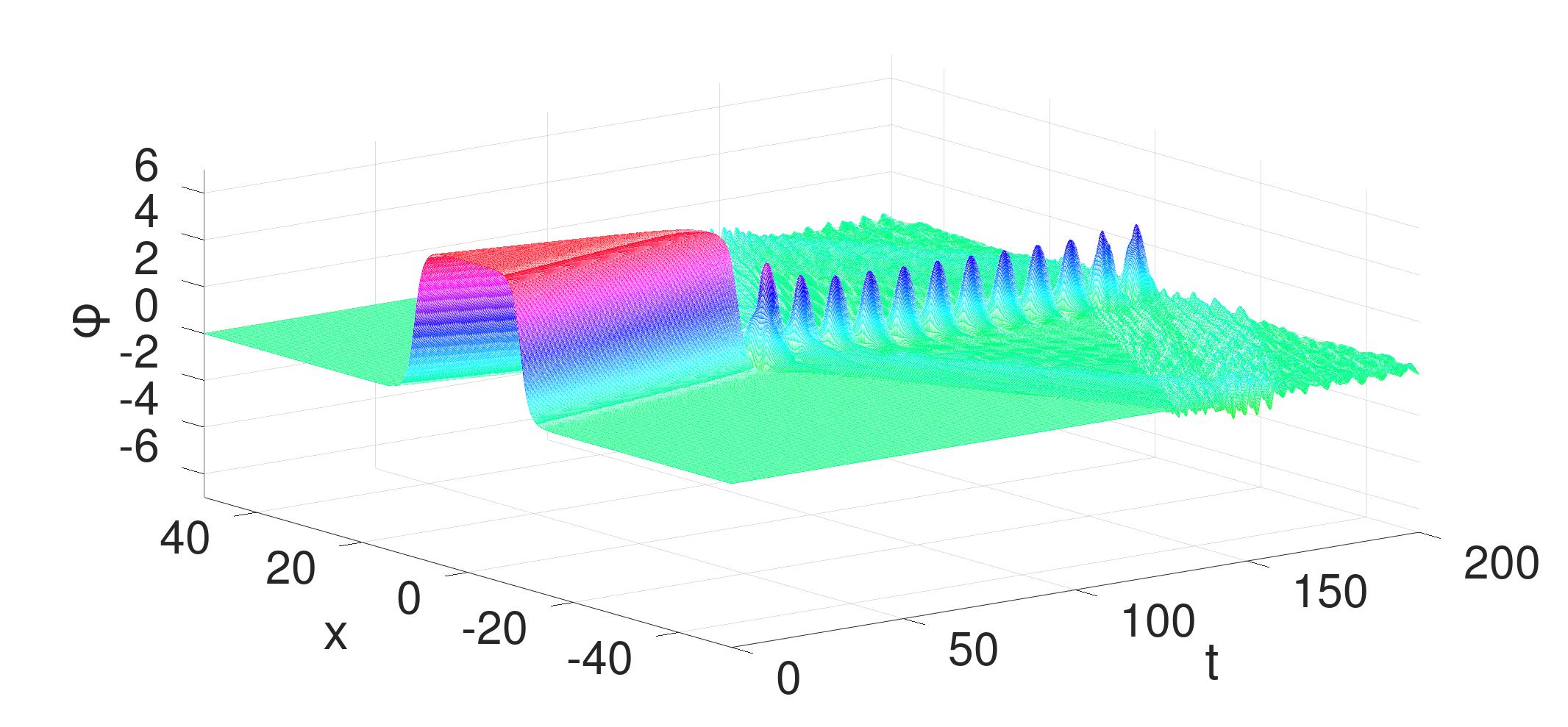} \label{fig:BionForm_a}}
	\quad
	\subfloat[$v_{\rm in }=0.201$ for $n=1$ and $\alpha = 1$.]
	{\includegraphics[scale=0.1]{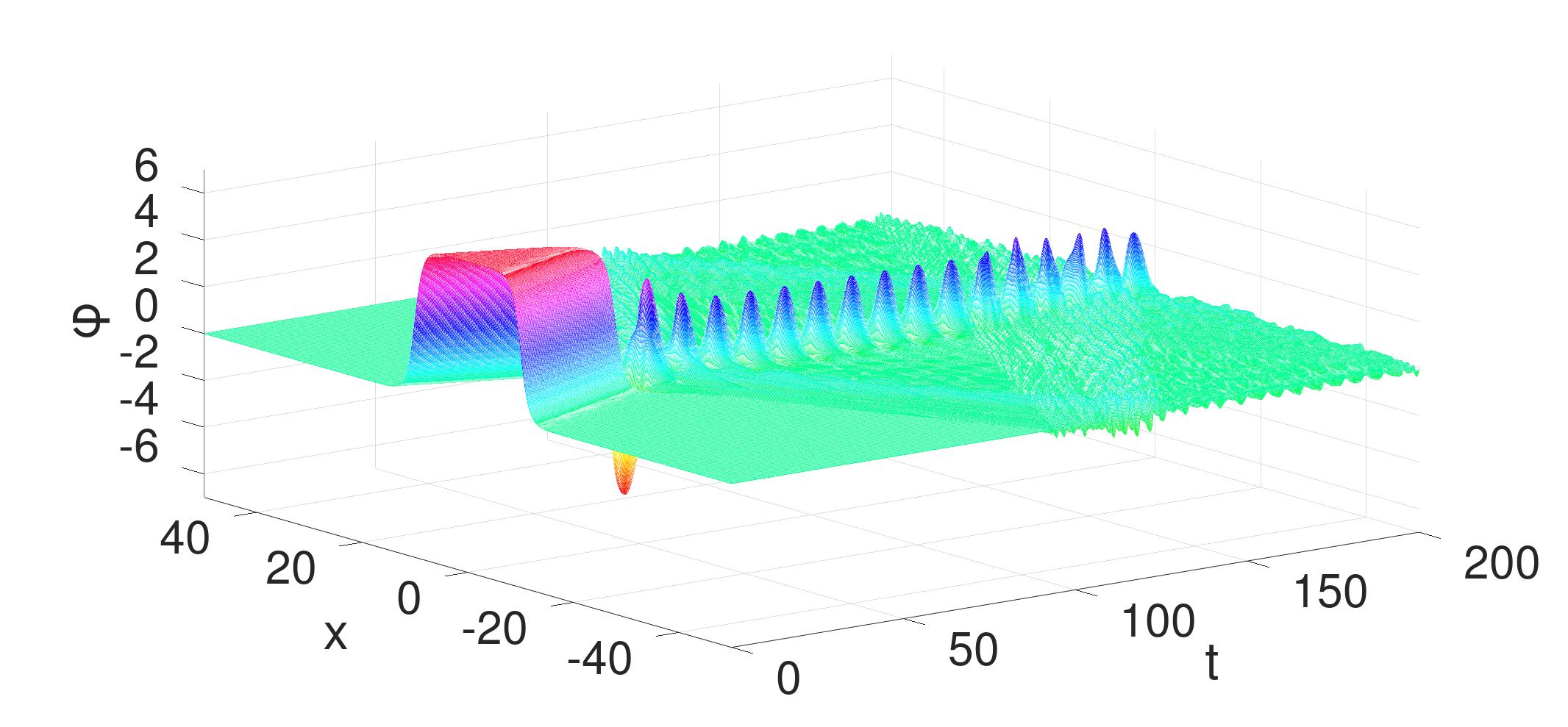} \label{fig:BionForm_b}}
	\quad 
	\subfloat[$v_{\rm in }=0.301$ for $n=2$ and $\alpha=1$.]
	{\includegraphics[scale=0.1]{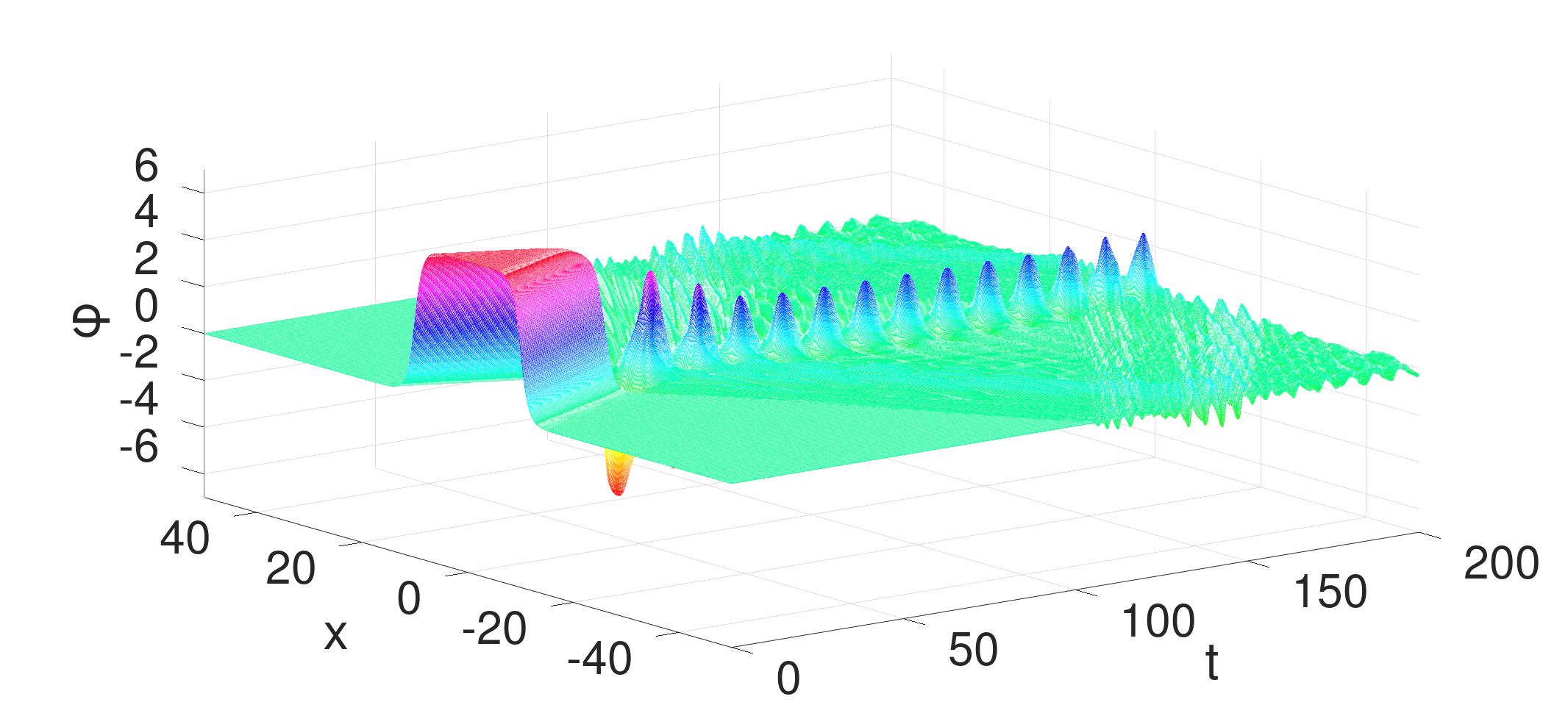} \label{fig:BionForm_c} }
	\caption{\label{fig:BionForm}Bion Formation}
\end{figure}
\begin{align}
	\varphi(x,0) & = \varphi_{K} \left(\frac{x+x_{0}}{\sqrt{1-v_{\rm in}^{2}}}\right) + \varphi_{\overline{K}} \left(\frac{x-x_{0}}{\sqrt{1-v_{\rm in}^{2}}}\right) - 2\pi \\
	\dot{\varphi}(x,0) & = -\frac{v_{\rm in}}{\sqrt{1-v_{\rm in}^{2}}} \varphi_{K}^{\prime}\left(\frac{x+x_{0}}{\sqrt{1-v_{\rm in}^{2}}}\right) - \frac{v_{\rm in}}{\sqrt{1-v_{\rm in}^{2}}} \varphi_{\overline{K}}^{\prime} \left(\frac{x-x_{0}}{\sqrt{1-v_{\rm in}^{2}}}\right)
\end{align}
where the primes denote the derivatives with respect to the argument, $v_{\rm in}$ is twice the relative velocity between the kink and antikink, and $x_{0}$ is half the separation of the kink and antikink, which we set equal to $10$ in our numerical computations.
\begin{figure}
	\centering
	\subfloat[$v_{\rm in}=0.591$ for $n=1$ and $\alpha = 5$.]
	{\includegraphics[scale=0.1]{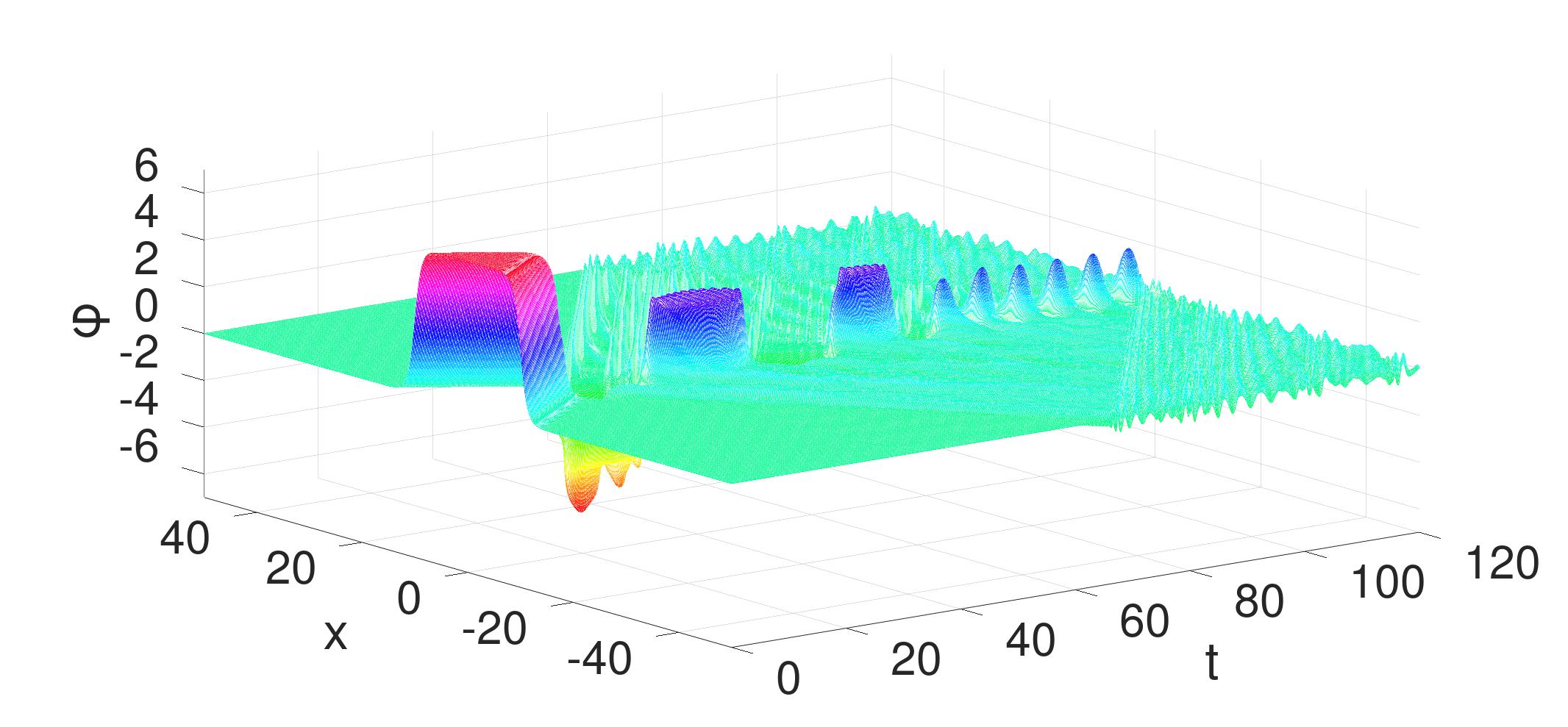} \label{fig:BionOrigin_a}}
	\quad
	\subfloat[$v_{\rm in }=0.599$ for $n=1$ and $\alpha = 5$.]
	{\includegraphics[scale=0.1]{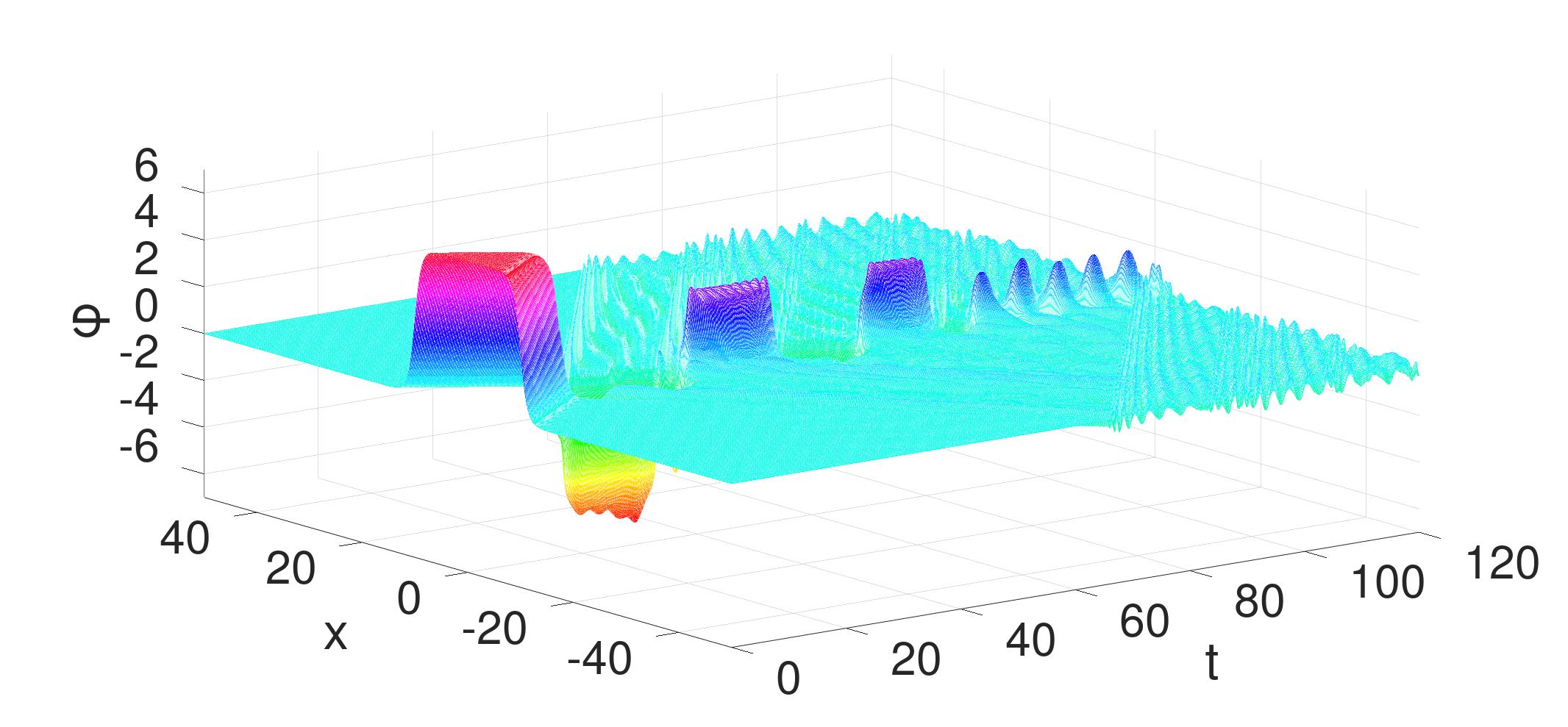}\label{fig:BionOrigin_b}}
	\quad
	\subfloat[$v_{\rm in }=0.553$ for $n=2$ and $\alpha=5$.]
	{\includegraphics[scale=0.1]{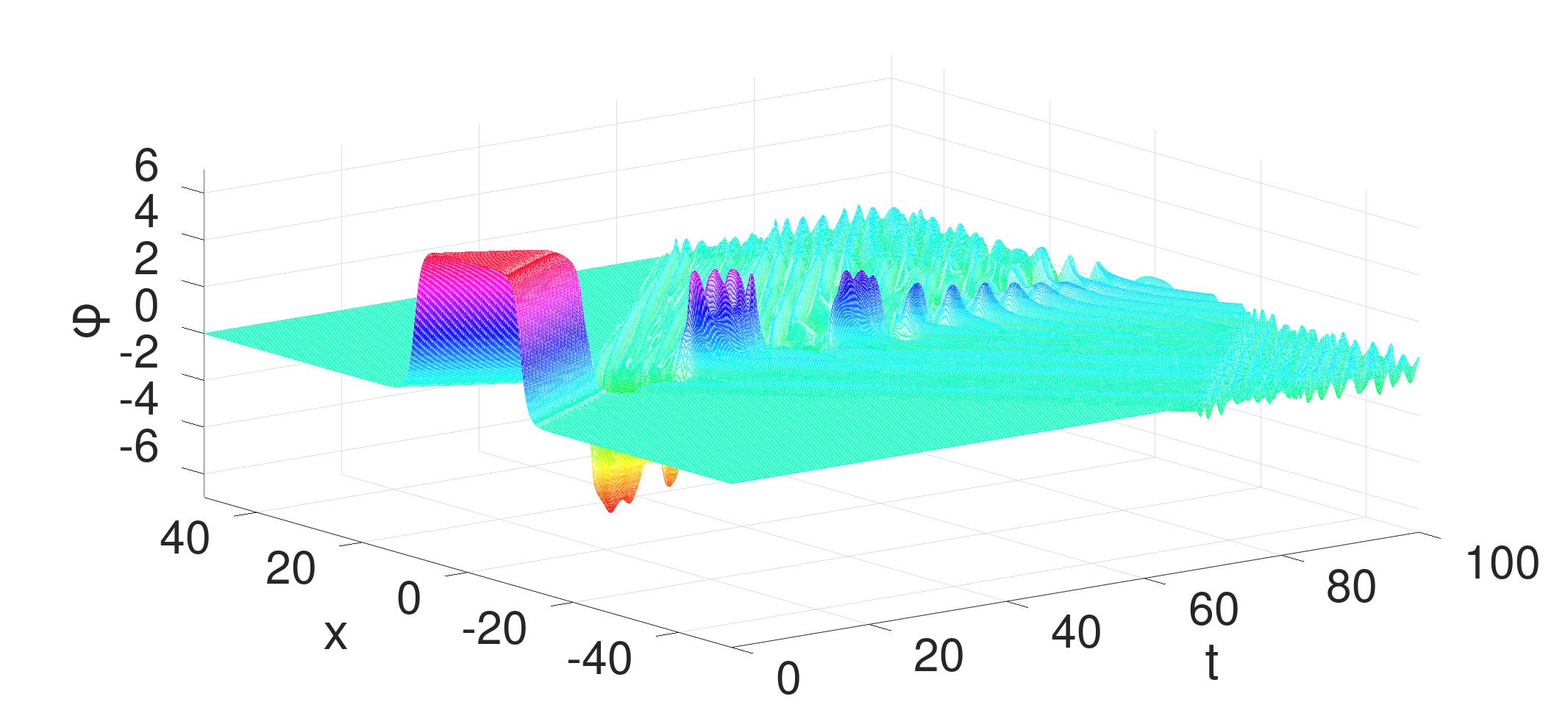}\label{fig:BionOrigin_c}}
	\quad 
	\subfloat[$v_{\rm in}=0.554$ for $n=2$ and $\alpha = 5$.]
	{\includegraphics[scale=0.1]{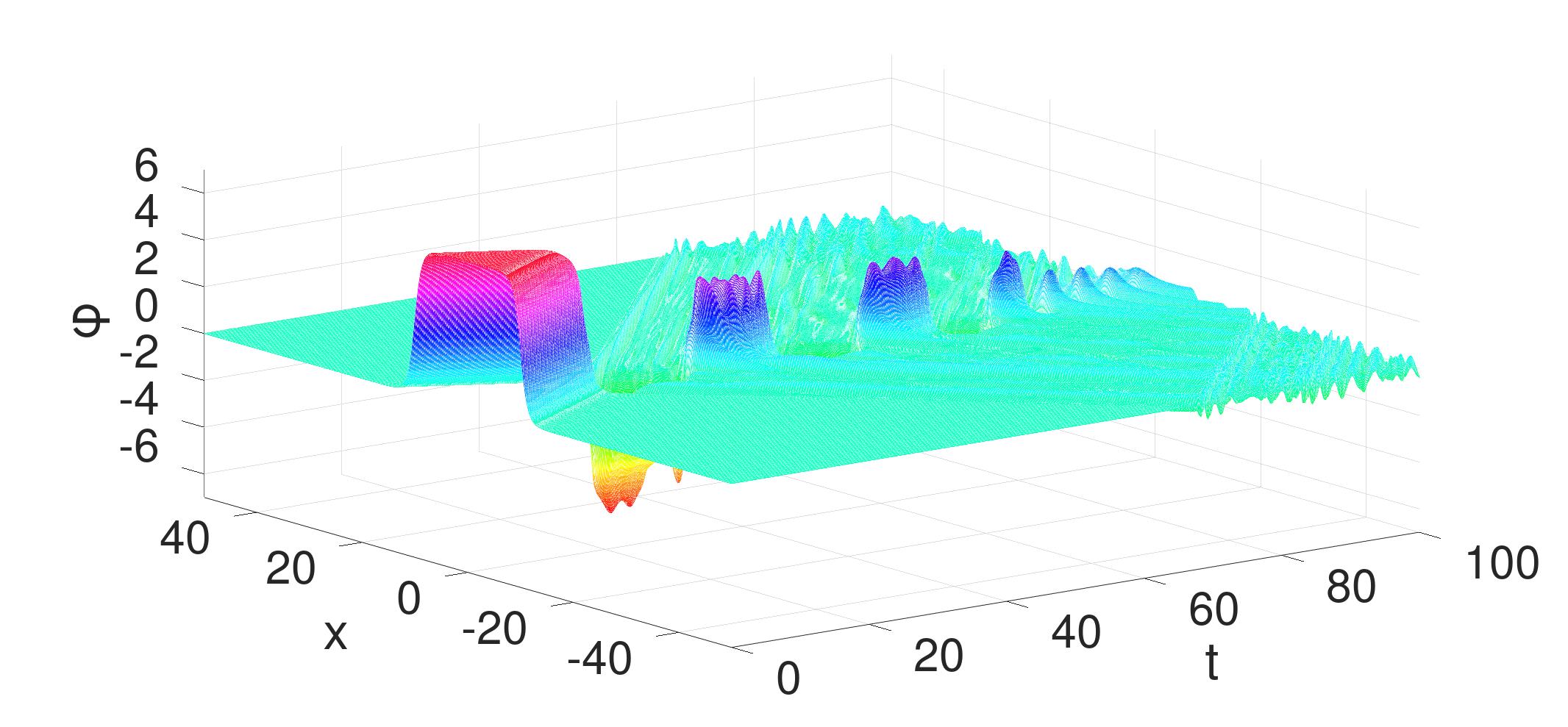} \label{fig:BionOrigin_d}}
	\quad 
	\subfloat[$v_{\rm in }=0.585$ for $n=1$ and $\alpha = 20$]
	{\includegraphics[scale=0.1]{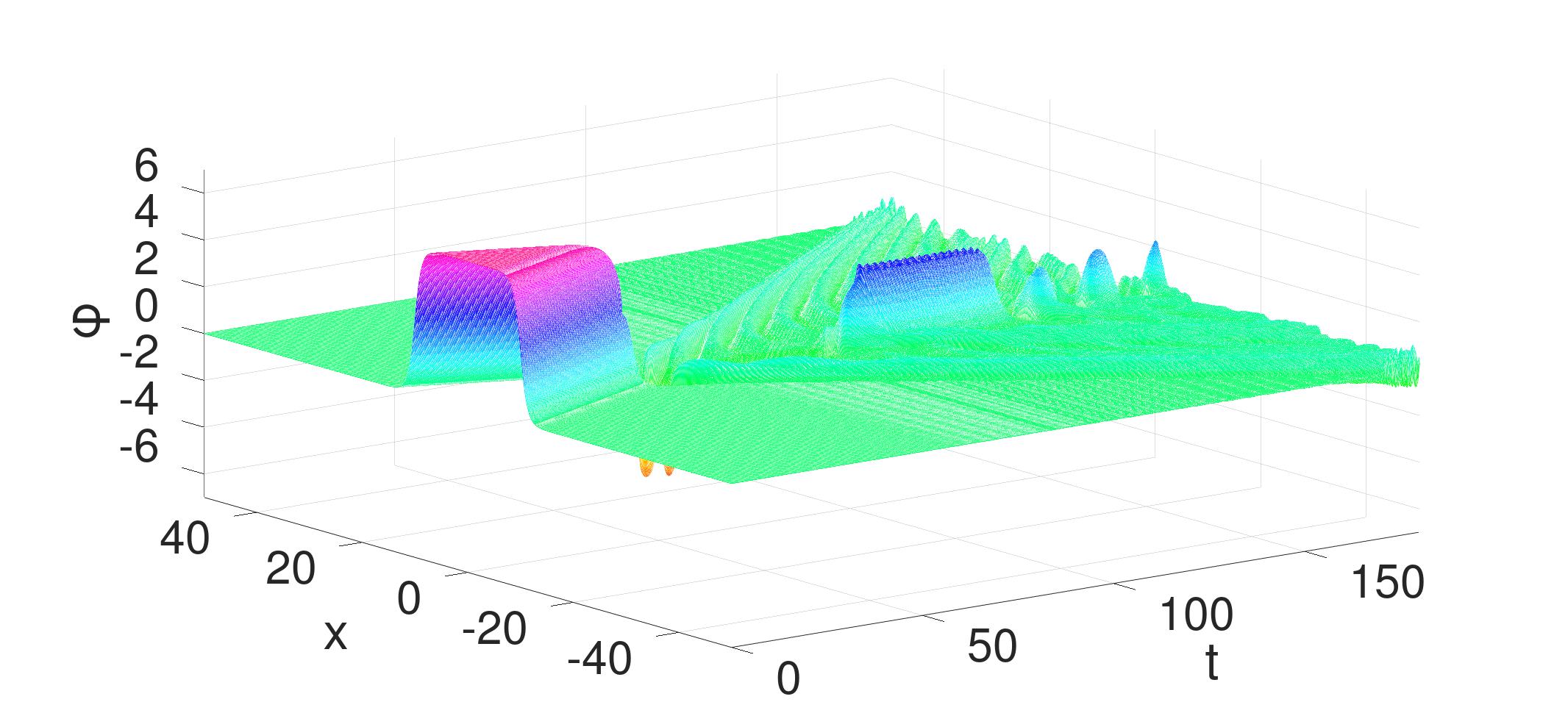} \label{fig:BionOrigin_e}}
	\quad 
	\subfloat[$v_{\rm in }=0.651$ for $n=2$ and $\alpha=20$]
	{\includegraphics[scale=0.1]{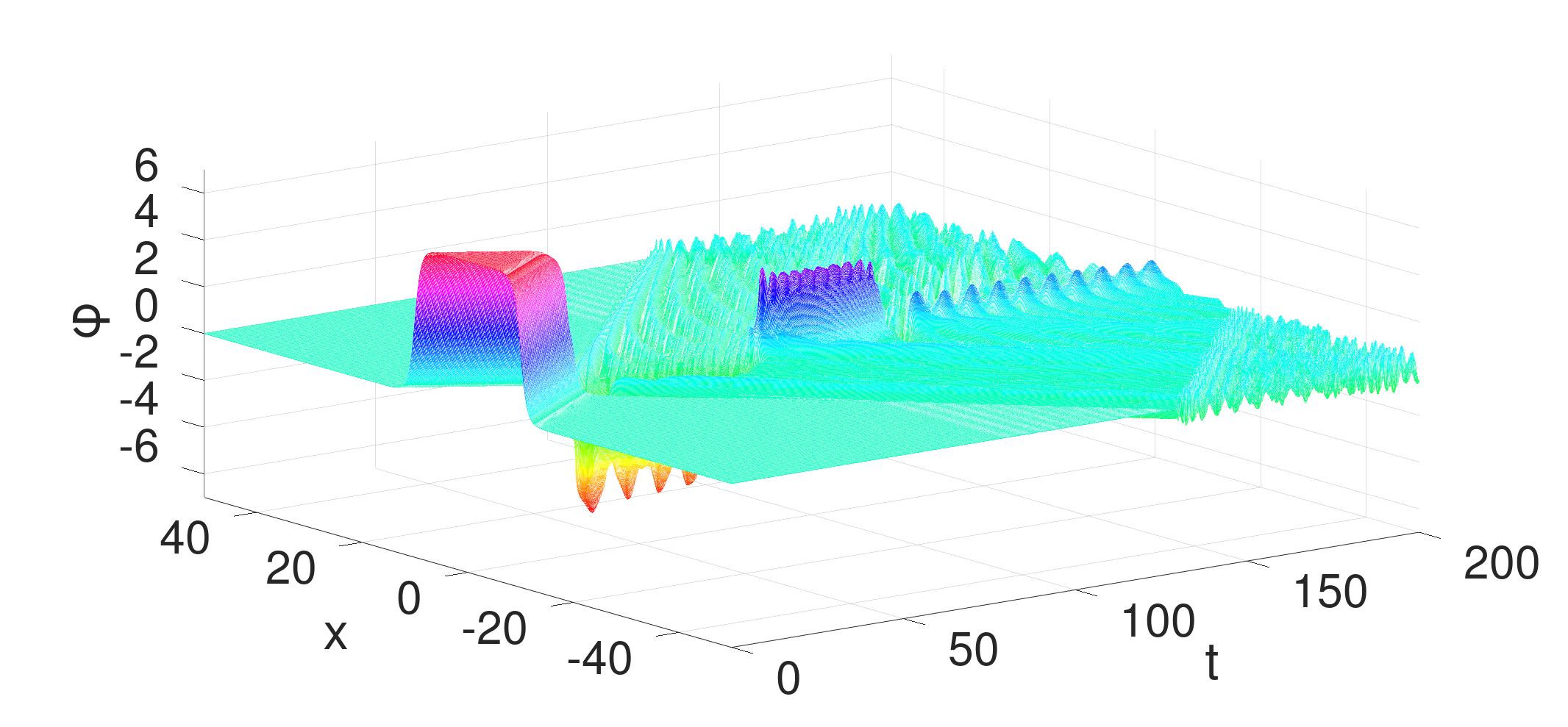}\label{fig:BionOrigin_f}}
	\caption{\label{fig:BionOrigin} Formation of pairs of bions at the origin of the kink.}
\end{figure}

We solve the dynamical equation as an initial value problem using the Fourier spectral method on a grid with $N=1800$ nodes with periodic boundary conditions for $x \in \left[-50,50\right]$. Discretization in $x$ yields a system of second-order ordinary differential equations in time, which are solved by integrating with $t$ using the ode45 routine of MATLAB. We set the tolerance option for the ode45 routine as ${\rm RelTol}=10^{-10}$ and ${\rm AbsTol=10^{-12}}$ to ensure that the total energy is conserved. The observed critical velocities for various values of $n$ and $\alpha$ are indicated in Table \ref{t3}. 

In the absence of the internal shape mode, thus for $\alpha<4$, we observe, in one case, a long-lived oscillating state after the annihilation of the kink and antikink pair. This is shown in Figure \ref{fig:reflection}. This occurs when the initial velocities $v_{\rm in} $ are greater than the critical velocities, $v_{\rm cr}$; as indicated in Figure \ref{fig:reflection_a} where $v_{\rm in} =0.501 > v_{\rm cr} =0.3626 $ and Figure \ref{fig:reflection_b} where $v_{\rm in} =0.400 > v_{\rm cr} =0.3998 $. 
Also, for some initial velocities less than the critical velocity, an annihilation of kink and antikink to a centrally-located bion occurs. These observed features are shown in Figure \ref{fig:AnnCenBion}, where in Figure \ref{fig:AnnCenBion_a} $v_{\rm in} =0.361 < v_{\rm cr} =0.3626 $, Figure \ref{fig:AnnCenBion_b} $v_{\rm in}= 0.3621 < v_{\rm cr} =0.3626 $ and in Figure \ref{fig:AnnCenBion_c} $v_{\rm in} =0.3961 < v_{\rm cr} =0.3998 $. Furthermore, for initial velocities less than the critical velocities, an annihilation of the kink and antikink pair to a long-living bound state occurs with radiation of energy in the form of small amplitude waves. This is indicated in Figure \ref{fig:BionForm_a} where $v_{\rm in} =0.101 < v_{\rm cr} =0.3626 $, Figure \ref{fig:BionForm_a} where $v_{\rm in} =0.201 < v_{\rm cr} =0.3626 $ and Figure \ref{fig:BionForm_c} where $v_{\rm in} =0.301 < v_{\rm cr} =0.3998 $. 

In the presence of the internal shape mode, thus for $\alpha \geq 4$, we observe that, for $v_{\rm in} < v_{\rm cr}$ after one-two bounces, the exchange energy of the kink and antikink is not enough to overcome the attractive potential as a result the exchange energy is converted to radiation leading to a bion state (see Figures \ref{fig:BionOrigin_a}, \ref{fig:BionOrigin_b}, \ref{fig:BionOrigin_c}, \ref{fig:BionOrigin_d}, \ref{fig:BionOrigin_e} and \ref{fig:BionOrigin_f} respectively). The bion remains at the origin of the kink in this case after the collision. Thus, the kink and antikink do not escape after the collision. The results show that there is a transfer of energy into the internal shape mode but once the shape modes are destroyed they are unable to overcome their initial interaction resulting in a bounce state after the first and second interaction. Also, for for $v_{\rm in} > v_{\rm cr}$ in this regime, we observe a long-lived oscillating state as reported in Figure \ref{fig:reflection_c} where $v_{\rm in} =0.601 > v_{\rm cr} =0.555 $.

\begin{figure}
	\centering
	\subfloat[The case for $n=1$ and $\alpha = 5$.]
	{\includegraphics[scale=0.3]{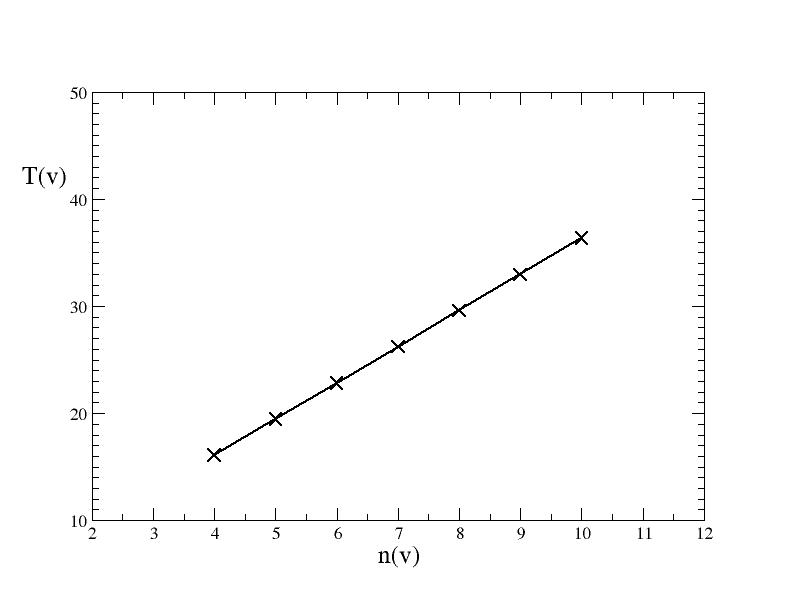} \label{fig:timebounce_a}}
	\quad
	\subfloat[The case for $n=2$ and $\alpha = 5$.]
	{\includegraphics[scale=0.3]{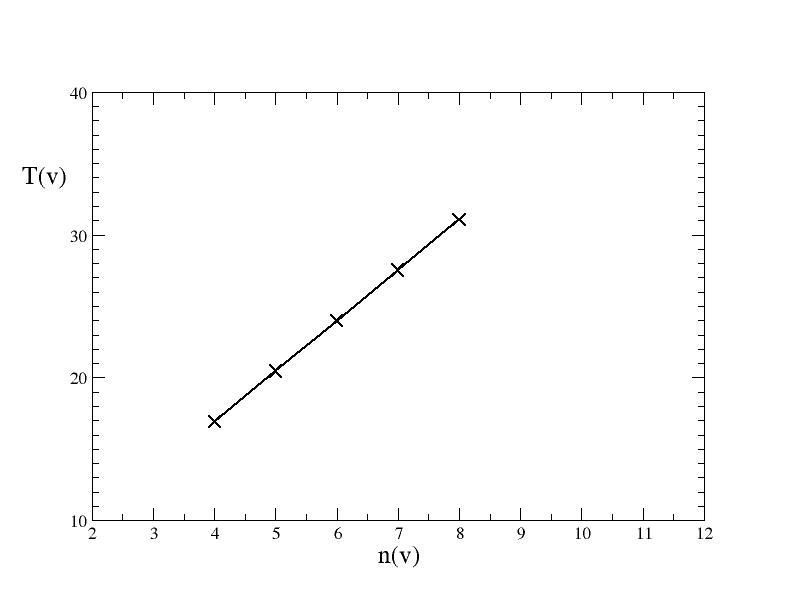}\label{fig:timebounce_b}}
	\caption{\label{fig:timebounce} The time between the two collisions in the two bounce windows verses the number of oscillations.}
\end{figure}

The points marked ``X" in Figures \ref{fig:timebounce_a} and \ref{fig:timebounce_b} represent the time $T(v)$ between the first and second collisions as a function of the number of oscillations between these collisions for $n = 1, \hspace{0.05cm} \alpha =5$ and $n = 2, \hspace{0.05cm} \alpha =5$, respectively.
The results fit the resonance mechanism relation proposed in~\cite{Campbell:1986mg,Campbell:1983xu}, with $\omega_{T} T(v) = 2\pi n(v) + \delta$ to some extent. The only difference is that $\omega_{T} \approx 2\omega_{s} $ where $\omega_{s}$ is chosen close to the internal shape mode for the considered free parameters. In this case, $\delta=4.832$ falls between $0$ and $2\pi$. Another inconsistency in this model is the relationship between $T(v)$ and the binding energy. The relationship is given by~\cite{Campbell:1986mg,Campbell:1983xu} $T(v) \propto c\left(v_{\rm cr}^{2} - v_{\rm in}^{2}\right)^{-\beta} $, where $v_{\rm in}$ denotes impact velocities less than the critical velocity $v_{\rm cr}$, with $c=1$ and $\beta=0.5$. In this model, the fit to this relation favors a smaller value of $\beta$, in the range of $0.2-0.357$ and $c=2$. The differences in this model may be due to the nature of the resonance mechanism: in this model, the energy remains within the kink and antikink for an extended period of time, resulting in a large amount of radiation after the first and second collisions, whereas in the resonant $\phi^{4}$ scattering~\cite{Campbell:1983xu}, the energy is rapidly localized on the kink and antikink. There were, however, similarities between this model and the $\phi^{4}$ model. Two bounce windows are missing for $n(v)<4$, as seen in the resonant $\phi^{4}$ model for $n(v)<3$. Figures \ref{fig:BionOrigin_a} and \ref{fig:BionOrigin_b} depict windows for $n(v)=8$ and $n(v)=9$ for $n=1$ and $\alpha=5$, whereas Figures \ref{fig:BionOrigin_c} and \ref{fig:BionOrigin_d} depict windows for $n(v)=7$ and $n(v)=8$ for $n=2$ and $\alpha=5$. 
\begin{table}
	\caption{\label{t3} Prediction of the critical velocities.}
	\centerline{
		\begin{tabular}{c|c|c}
			$n$ & $\alpha$ & $v_{\rm cr}$ \\
			\hline
			1   &   1      & 0.3626 \\
			1   &   5      & 0.601 \\
			1   &   10     & 0.669 \\ 
			1   &   20     & 0.701 \\ 
			2   &   1      & 0.3998 \\
			2   &   5      & 0.555 \\
			2   &   10     & 0.617 \\
			2   &   20     & 0.698  
		\end{tabular}
	}
\end{table}

\section{Conclusion}
\label{sec:concl}
We have presented the numerical simulation of the scattering of kinks in the noncanonical sine-Gordon model. This model contains two free parameters, $\alpha$ and $n$. In the limit $\alpha=0$, we recover the sine-Gordon model. Central to this analyzes has been the presence of (localized) inner structures in the energy density of the kinks. These inner structures arise by varying the free parameters that make up the model for $\alpha \geq 1$. We anticipated in our numerical analyzes that the presence of these localized structures in the energy-density of this model will yield resonance structures and bion-formations in the scattering of the kink and antikink.

In analyzing the linear spectrum of the static kinks for $n=1, 2$ and $\alpha \geq 0$, we observe that both the translational and internal shape modes contribute. The shape modes are only observed for $\alpha \geq 4$. The energy-eigenvalues of the shape mode for $n=1,2$ decrease monotonically as $\alpha$ increases. From analyzes of the study by the authors in Refs.~\cite{Campbell:1983xu,Campbell:1986mg} it was found that the shape mode plays a crucial role in the formation of resonance structures in the scattering of the kink and antikink. They argued that the formation of the resonance structures is a result of the transfer of energy from the shape mode to the translational zero mode. However, it was later discovered by the authors in Ref.~\cite{Dorey:2011yw} that in the absence of shape mode resonance features occur. Thus, a robust explanation of the formation of the resonance structures is needed. Recent studies from Ref.~\cite{Zhong:2019fub} reveal that the presence of localized inner structures also plays an important role in the formation of the resonance structures.

In our numerical analysis, we discovered that as $n$ increases from one to four for $\alpha = 1$, the energy density of the static kink splits into two to three peaks. The separation between these peaks becomes wider and shallower as $\alpha$ increases. We found that in the absence of the internal shape mode, no resonance structures are observed in the scattering of the kink and antikink for various values of $\alpha$ for $n=1$ and $n=2$. We only observed three bion features: a long-lived oscillating state with small amplitudes; a large amplitude bion formation; and annihilation of a kink and antikink to a centrally-located bion. However, two-bounce bion formation is observed in the presence of the shape mode, thus $\alpha \geq 4$. The latter observation may be attributed to the broad and shallow nature of the peaks of the kink and antikink energy density.

To conclude, our results show bion-formations (static and moving oscillations) in the kink-antikink scattering for the free parameters $\alpha$ and $n$. It will also be interesting to look at the behavior of the kink and antikink collision in other noncanonical sine-Gordon models that possess inner structures in their energy density, like the one proposed by the authors in Ref.~\cite{Zhong:2014kha}. Other interesting avenues to consider are the $\phi^{6}$ type superpotential and those that yield double kink solutions. We are currently considering these for our future work.

\section*{Acknowledgements}
We are grateful to Prof. H. Weigel for reading the manuscript and for helpful comments.


\begin{thebibliography}{99}
\bibitem{Rajaraman:1982is} R.~Rajaraman. Solitons and Instantons. North Holland, 1982.

\bibitem{Vachaspati:2006zz} T.~Vachaspati. Kinks and domain walls: An introduction to classical and quantum solitons. Cambridge University Press, 2010.

\bibitem{Vilenkin:2000jqa} A.~Vilenkin and E.~P.~S.~Shellard. Cosmic Strings and Other Topological Defects. Cambridge University Press, 2000.

\bibitem{Kevre2008} P.~G.~Kevrekidis, D.~J.~Frantzeskakis, and R.~Carretero-González. Emergent Nonlinear Phenomena in Bose-Einstein Condensates: Theory and Experiment. Springer-Verlag, 2008.

\bibitem{Ivanov:1992aa} B.~Ivanov, A.~Kichiziev, and Y.~N.~Mitsai. Nonlinear Dynamics and Relaxation of Strongly Anisotropic Ferromagnets. Soviet Physics JETP 1992; 75(2): 329--337. 

\bibitem{bishop1980solitons} A.~R.~Bishop, J.~A.~Krumhansl, and S.~E.~ Trullinger. Solitons in Condensed Matter: A Paradigm. Physica D 1980; 1(1): 1--44. doi: 10.1016/0167-2789(80)90003-2

\bibitem{Weigel:2008zz} H.~Weigel. Chiral soliton models for baryons. (Vol. 743), Springer-Verlag, 2008.

\bibitem{Takyi:2019ahv} I.~Takyi and H.~Weigel. Nucleon Structure Functions from the NJL-Model Chiral Soliton. Europen Physics Journal A 2019; 55(8): 128. doi: 10.1140/epja/i2019-12806-3

\bibitem{Weigel:2021pbr} H.~Weigel and I.~Takyi. Chiral Soliton Models and Nucleon Structure Functions. Symmetry 2021; 13(1): 108. doi: 10.3390/sym13010108

\bibitem{Campbell:1986mg} D.~K.~Campbell and M.~Peyrard. Solitary Wave Collisions Revisited. Physica D 1986; 18: 47--53. doi: 10.1016/0167-2789(86)90161-2

\bibitem{Campbell:1983xu} D.~K.~Campbell, J.~F.~Schonfeld, and C.~A.~Wingate. Resonance Structure in Kink - Antikink Interactions in $\phi^{4}$ Theory. Physica D 1983; 9: 1. doi: 10.1016/0167-2789(83)90289-0

\bibitem{Belova:1997bq} T.~I.~Belova and A.~E.~Kudryavtsev. Solitons and their Interactions in Classical Field Theory. Physics-Uspekhi 1997; 40: 359--386. doi: 10.1070/PU1997v040n04ABEH000227 

\bibitem{Anninos:1991un} P.~Anninos, S.~Oliveira, and R.~A.~Matzner. Fractal Structure in the Scalar $\lambda (\phi^2-1)^2$ Theory. Physics Review D 1991; 44: 1147--1160. doi: 10.1103/PhysRevD.44.1147

\bibitem{Ablowitz:1979a} M.~J.~Ablowitz, M.~D.~Kruskal, and J.~F.~Ladik. Solitary Wave Collisions. SIAM Journal on Applied Mathematics 1979; 36(3): 428--437. doi: 10.1137/0136033

\bibitem{Moshir:1981ja} M.~Moshir. Soliton - Anti-soliton Scattering and Capture in $\lambda \phi^4$ Theory. Nuclear Physics B 1981; 185: 318--332. doi: 10.1016/0550-3213(81)90320-5

\bibitem{Goodman:2005ja} R.~Goodman and R.~Haberman. Kink-Antikink Collisions in the $\phi^4$ Equation: The n-Bounce Resonance and the Separatrix Map. SIAM Journal on Applied Dynamical Systems 2005; 4(4): 1195--1228. doi: 10.1137/050632981

\bibitem{Goodman:2007e} R.~H.~Goodman and R.~Haberman. Chaotic scattering and the n-bounce resonance in solitary-wave interactions. Physics Review Letters 2007; 98(10): 104103. doi: 10.1103/PhysRevLett.98.104103

\bibitem{Weigel:2013kwa} H.~Weigel. Kink-Antikink Scattering in $\varphi^4$ and $\phi^6$ Models. Journal of Physics: Conference Series 2014; 482: 012045. doi: 10.1088/1742-6596/482/1/012045

\bibitem{Dorey:2011yw} P.~Dorey, K.~Mersh, T.~Romanczukiewicz, and Y.~Shnir. Kink-Antikink Collisions in the $\phi^6$ Model. Physics Review Letters 2011; 107: 091602. doi: 10.1103/PhysRevLett.107.091602

\bibitem{Belendryasova:2019eq} E.~Belendryasova and V.~A.~Gani. Scattering of the $\varphi^8$ Kinks with Power-Law Asymptotics. Communications in Nonlinear Science and Numerical Simulation 2019; 67: 414--426. doi: 10.1016/j.cnsns.2018.07.030 

\bibitem{Campos:2020ust} J.~G.~F.~Campos and A.~Mohammadi. Interaction Between Kinks and Antikinks with Double Long-Range Tails. Physics Letters B 2021; 818: 136361. doi: 10.1016/j.physletb.2021.136361

\bibitem{Gani:2015cda} V.~A.~Gani, V.~Lensky, and M.~A.~Lizunova. Kink Excitation Spectra in the (1+1)-Dimensional $\varphi^8$ Model. Journal of High Energy Physics 2015; 08: 147. doi: 10.1007/JHEP08(2015)147

\bibitem{Christov:2018ecz} I.~C.~Christov, R.~J.~Decker, A.~Demirkaya, V.~A.~Gani, P.~G.~Kevrekidis, A.~Khare, and A.~Saxena. Kink-Kink and Kink-Antikink Interactions with Long-Range Tails. Physics Review Letters 2019; 122(17): 171601. doi: 10.1103/PhysRevLett.122.171601

\bibitem{Christov:2018wsa} I.~C.~Christov, R.~J.~Decker, A.~Demirkaya, V.~A.~Gani, P.~G.~Kevrekidis, and R.~V.~Radomskiy. Long-Range Interactions of Kinks. Physics Review D 2019; 99(1): 016010. doi: 10.1103/PhysRevD.99.016010

\bibitem{Manton:2018deu} N.~S.~Manton. Forces Between Kinks and Antikinks with Long-Range Tails. Journal of Physics A 2019; 52(6): 065401. doi: 10.1088/1751-8121/aaf9d1

\bibitem{Bazeia:2018bhf} D.~Bazeia, R.~Menezes, and D.~C.~Moreira. Analytical Study of Kinklike Structures with Polynomial Tails. Journal of Physics Communications 2018; 2(5): 055019. doi: 10.1088/2399-6528/aac3cd

\bibitem{Gani:2020pio} V.~A.~Gani, A.~M.~Marjaneh, and P.~A.~Blinov. Explicit Kinks in Higher-Order Field Theories. Physics Review D 2020; 101(12): 125017. doi: 10.1103/PhysRevD.101.125017

\bibitem{Christov:2020zhb} I.~C.~Christov, R.~J.~Decker, A.~Demirkaya, V.~A.~Gani, P.~G.~Kevrekidis, and A.~Saxena. Kink-Antikink Collisions and Multi-Bounce Resonance Windows in Higher-Order Field Theories.  Communications in Nonlinear Science and Numerical Simulations 2021; 97: 105748. doi: 10.1016/j.cnsns.2021.105748

\bibitem{Gani:2014gxa} V.~A.~Gani, A.~E.~Kudryavtsev, and M.~A.~Lizunova. Kink Interactions in the (1+1)-Dimensional $\phi^6$ Model. Physics Review D 2014; 89(12): 125009. doi: 10.1103/PhysRevD.89.125009

\bibitem{Bazeia:2017rxo} D.~Bazeia, E.~Belendryasova and V.~A.~Gani. Scattering of kinks of the sinh-deformed $\varphi^4$ model. European Physics Journal C 2018; 78(4): 340. doi: 10.1140/epjc/s10052-018-5815-z

\bibitem{Bazeia:2019xoe} D.~Bazeia, A.~R.~Gomes, K.~Z.~Nobrega, F.~C.~Simas. Kink scattering in hyperbolic models. International Journal Modern Physics A 2019; 34(31): 1950200. doi: 10.1142/S0217751X19502002

\bibitem{Takyi:2020nkn} I.~Takyi, B.~Barnes, J.~Ackora-Prah. Vacuum Polarization Energy of the Kinks in the Sinh-Deformed Models. Turkish Journal of Physics 2021; 45: 194--206. doi: doi:10.3906/fiz-2103-32

\bibitem{Takyi:2016tnc} I.~Takyi and H.~Weigel. Collective Coordinates in One-Dimensional Soliton Models Revisited. Physics Review D 2016; 94(8): 085008. doi: 10.1103/PhysRevD.94.085008

\bibitem{Christov:2008kk} I.~Christov and I.~C~Christov. Physical Dynamics of Quasi-Particles in Nonlinear Wave Equations. Physics Letters A 2008; 372(6): 841--848. doi: S0375960107012327

\bibitem{Manton2004} N.~Manton and P.~Sutcliffe. Topological solitons. Cambridge University Press, 2004. doi: 10.1017/CBO9780511617034

\bibitem{Manton:1978gf} N.~S.~Manton. An Effective Lagrangian for Solitons. Nuclear Physics B 1979; 150: 397--412. doi: 10.1016/0550-3213(79)90309-2

\bibitem{Manton:2021ipk} N.~S.~Manton, K.~Oles, T.~Romanczukiewicz, and A.~Wereszczynski. Collective Coordinate Model of Kink-Antikink Collisions in $\phi^4$ Theory. Physics Review Letters 2021; 127(7): 071601. doi: 10.1103/PhysRevLett.127.071601

\bibitem{Kevrekidis:2004ga} P.~G.~Kevrekidis, A.~Khare, and A.~Saxena. Solitary Wave Interactions in Dispersive Equations Using Manton's Approach. Physics Review E 2004; 70(5): 057603. doi: 10.1103/PhysRevE.70.057603

\bibitem{Goodman:2004ef} R.~H.~Goodman and R.~Haberman. Interaction of sine-Gordon Kinks with Defects: the Two-Bounce Resonance. Physica D 2004; 195(3): 303--323. doi: 10.1016/j.physd.2004.04.002

\bibitem{Fei:1992aj} Z.~Fei, Y.~S.~Kivshar, L.~Vázquez. Resonant Kink-Impurity Interactions in the sine-Gordon Model. Physics Review A 1992; 45(8): 6019--6030. doi: 10.1103/PhysRevA.45.6019

\bibitem{Kivshar:1991zz} Y.~S.~Kivshar, Z.~Fei, and L.~Vasquez. Resonant Soliton-Impurity Interactions. Physics Review Letters 1991; 67: 1177--1180. doi: 10.1103/PhysRevLett.67.1177

\bibitem{Zhang:1991ee} F.~Zhang, Y.~S.~Kivshar, B.~A.~Malomed, and L.~Vázquez. Kink Capture by a Local Impurity in the sine-Gordon Model. Physics Letters A 1991; 159(6): 318--322. doi: 10.1016/0375-9601(91)90440-J

\bibitem{Dorey:2021mdh} P.~Dorey, A.~Gorina, I.~Perapechka, T.~Roma\'nczukiewicz, and Y.~Shnir. Resonance Structures in Kink-Antikink Collisions in a Deformed sine-Gordon Model. Journal of High Energy Physics 2021; 145: 9. doi: 10.1007/JHEP09(2021)145

\bibitem{Gani:2017yla} V.~A.~Gani, A.~M.~Marjaneh, A.~Askari, E.~Belendryasova, and D.~Saadatmand. Scattering of the Double sine-Gordon Kinks. European Physics Journal C 2018; 78(4): 345. doi: 10.1140/epjc/s10052-018-5813-1

\bibitem{Gani:1998jb} V.~A.~Gani and A.~E.~Kudryavtsev. Kink - Antikink Interactions in the Double sine-Gordon Equation and the Problem of Resonance Frequencies. Physics Review E 1999; 60: 3305--3309. doi: 10.1103/PhysRevE.60.3305

\bibitem{Campbell:1986nu} D.~K.~Campbell, M.~Peyrard, and P.~Sodano. Kink - Antikink Interactions in the Double {Sine-Gordon} Equation. Physica D 1986; 19: 165--205. doi: 10.1016/0167-2789(86)90019-9

\bibitem{Malomed:1989gx} B.~A.~Malomed. Dynamics and Kinetics of Solitons in the Driven Damped Double sine-Gordon Equation. Physics Letters A 1989; 136: 395--401. doi: 10.1016/0375-9601(89)90422-2

\bibitem{Nazifkar:2010aa} S.~Nazifkar and K.~Javidan. Collective Coordinate Analysis for Double sine-Gordon Model. Brazilian Journal of Physics 2010; 40: 102--107. doi: 10.1590/S0103-97332010000100014 

\bibitem{Zhong:2019fub} Y.~Zhong, X.~L.~Du, Z.~C.~Jiang, Y.~X.~Liu, and Y.~Q.~Wang.  Journal of High Energy Physics 2020; 02: 153. doi: 10.1007/JHEP02(2020)153

\bibitem{Bazeia:2007df} D.~Bazeia, L.~Losano, R.~Menezes, and J.~C.~R.~E.~Oliveira. Generalized Global Defect Solutions. European Physics Journal C 2007; 51: 953--962. doi: 10.1140/epjc/s10052-007-0329-0

\bibitem{Bazeia:2014dva} D.~Bazeia, A.~S.~Lobao, and R.~Menezes. Stable Static Structures in Models with Higher-Order Derivatives. Annals of Physics 2015; 360: 194--206. doi: 10.1016/j.aop.2015.05.017

\bibitem{Zhong:2018tbi} Y.~Zhong, R.~Z.~Guo, C.~E.~Fu, and Y.~X.~Liu. Kinks in Higher Derivative Scalar Field Theory. Physics Letters B 2018; 782: 346--352. doi: 10.1016/j.physletb.2018.05.048

\bibitem{Bazeia:2008tj} D.~Bazeia, L.~Losano, and R.~Menezes. First-Order Framework and Generalized Global Defect Solutions. Physics Letters B 2008; 668: 246--252. doi: 10.1016/j.physletb.2008.08.046

\bibitem{Zhong:2014kha} Y.~Zhong and Y.~X.~Liu. $K$-Field Kinks: Stability, Exact Solutions and New Features. Journal of High Energy Physics 2014; 10: 041. doi: 10.1007/JHEP10(2014)041
\end{thebibliography}

\end{document}